\documentclass[12pt,a4wide,epsf]{article}
\usepackage{epsfig,epsf}
\usepackage{epsfig}
\usepackage{graphicx}
\usepackage[german, english]{babel}
\usepackage{a4wide}
\usepackage{amsmath}
\usepackage{amssymb}
\usepackage{ifthen}
\usepackage{epsfig}

\newcounter{fig}    
\newcounter{fixy}

\begin{document}

\newcommand{\dd}{\mbox{d}}
\newcommand{\tr}{\mbox{tr}}
\newcommand{\la}{\lambda}
\newcommand{\ta}{\theta}
\newcommand{\f}{\phi}
\newcommand{\vf}{\varphi}
\newcommand{\ka}{\kappa}
\newcommand{\al}{\alpha}
\newcommand{\ga}{\gamma}
\newcommand{\de}{\delta}
\newcommand{\si}{\sigma}
\newcommand{\bomega}{\mbox{\boldmath $\omega$}}
\newcommand{\bsi}{\mbox{\boldmath $\sigma$}}
\newcommand{\bchi}{\mbox{\boldmath $\chi$}}
\newcommand{\bal}{\mbox{\boldmath $\alpha$}}
\newcommand{\bpsi}{\mbox{\boldmath $\psi$}}
\newcommand{\brho}{\mbox{\boldmath $\varrho$}}
\newcommand{\beps}{\mbox{\boldmath $\varepsilon$}}
\newcommand{\bxi}{\mbox{\boldmath $\xi$}}
\newcommand{\bbeta}{\mbox{\boldmath $\beta$}}
\newcommand{\ee}{\end{equation}}
\newcommand{\eea}{\end{eqnarray}}
\newcommand{\be}{\begin{equation}}
\newcommand{\bea}{\begin{eqnarray}}
\newcommand{\ii}{\mbox{i}}
\newcommand{\e}{\mbox{e}}
\newcommand{\pa}{\partial}
\newcommand{\Om}{\Omega}
\newcommand{\vep}{\varepsilon}
\newcommand{\bfph}{{\bf \phi}}
\newcommand{\lm}{\lambda}
\def\theequation{\arabic{equation}} 
\renewcommand{\thefootnote}{\fnsymbol{footnote}}
\newcommand{\re}[1]{(\ref{#1})}
\newcommand{\R}{{\rm I \hspace{-0.52ex} R}}
\newcommand{\N}{{\sf N\hspace*{-1.0ex}\rule{0.15ex}%
{1.3ex}\hspace*{1.0ex}}}
\newcommand{\Q}{{\sf Q\hspace*{-1.1ex}\rule{0.15ex}%
{1.5ex}\hspace*{1.1ex}}}
\newcommand{\C}{{\sf C\hspace*{-0.9ex}\rule{0.15ex}%
{1.3ex}\hspace*{0.9ex}}}
\newcommand{\eins}{1\hspace{-0.56ex}{\rm I}}
\renewcommand{\thefootnote}{\arabic{footnote}}

\title{Particle-like solutions to the Yang--Mills-dilaton \\system
in $d=4+1$ dimensions}

\author{{\large Eugen Radu}$^{\dagger}$, {\large Ya. Shnir}$^{\ddagger}$
 and {\large D. H. Tchrakian}$^{\dagger\star}$ \\ \\
$^{\dagger}${\small Department of Mathematical Physics, National
University of Ireland Maynooth,} \\ {\small Maynooth, Ireland}
\\ $^{\ddagger}${\small Institut f\"ur Physik, Universit\"at Oldenburg,
Postfach 2503 D-26111 Oldenburg, Germany}
\\ $^{\star}${\small School of Theoretical Physics -- DIAS, 10 
Burlington Road, Dublin 4, Ireland }}

\maketitle

\begin{abstract}
We construct static solutions to a  $SU(2)$ Yang--Mills (YM) dilaton model
in $4+1$ dimensions subject to bi-azimuthal symmetry.
The YM sector of the model consists of
the usual YM term and the next higher order term of the YM
hierarchy, which is required by the scaling condition for the existence
of finite energy solutions.
The basic features of two different types of configurations are studied, 
corresponding to (multi)solitons
with topological charge $n^2$, and soliton--antisoliton pairs
with zero topological charge. 
\end{abstract}

\section{Introduction}
Multi-instantons and composite instanton-antiinstanton bound states
subject to bi-azimuthal symmetry were reported in a recent
paper~\cite{Radu:2006gg}. These were
constructed numerically, for the usual ($p=1$) $SU(2)$ Yang-Mills (YM) system
in $4$ Euclidean dimensions, the spherically symmetric special case being the
usual BPST~\cite{BPST} instanton.

In a work~\cite{Brihaye:2002jg} unrelated to \cite{Radu:2006gg}, regular and
black hole static and spherically symmetric solutions
to a Einstein--YM (EYM) system in $4+1$ dimensional spacetime were constructed
numerically. The YM system in that model had gauge group $SO(4)$, with the
connection taking its values in (one of the two) chiral spinor
representations of $SO(4)$, namely in $SU(2)$. 
Given that the solutions in \cite{Brihaye:2002jg} were static, $i.e.$
that the YM field is defined on a $4$ dimensional Euclidean
space, the $SU(2)$ YM field in \cite{Brihaye:2002jg} 
is the same one as that in \cite{Radu:2006gg}.
This is the relation between the two works \cite{Radu:2006gg} and
\cite{Brihaye:2002jg}, and our intention here is to exploit this relation.

The present work serves two distinct purposes. The first and main purpose is
to pave the way for the construction of more general, non-spherically symmetric 
solutions to EYM systems in a five dimensional
spacetime. To our knowledge, no such results in EYM theory
have appeared in the literature
to date. Although considerable progress has been made in constructing
asymptotically flat higher dimensional EYM solutions\footnote{
EYM particle-like and black hole solutions 
approaching
at infinity the Minkowski 
background have been constructed numerically for $d=6,7$ and
$8$ \cite{Brihaye:2002hr}.
The properties of globally regular solutions in arbitrary
dimensions have been
studied both numerically and analytically in \cite{Breitenlohner:2005hx}.
Higher dimensional asymptotically
(anti-)de Sitter solutions have been found numerically  in
\cite{Radu:2005mj}, \cite{Brihaye:2006xc},
as well as such systems whose gravitational sector consists of higher
order Gauss-Bonnet like gravitational terms~\cite{Radu:2006mb}.}
all known configurations were
subject to spherical symmetry. Our choice of a YM-dilaton (YMd) model is made
because it has been shown that, at least in $d=3+1$ dimensions, 
the classical solutions of this system
mimic the corresponding EYM solutions~\cite{Maison:2004hb}, so
the dilaton--YM exercise serves as a
warmup for the considerably more complex gravitational problem.
 Our choice of a YMd model is made as an expedient in attempting the analysis
of the corresponding EYM model, the last being of physical interest low energy effective
actions of string theory, descended from $11$ dimensional supergravity. It is also a coincidence
here, that these supergravity descended low energy effective actions include the dilaton in
addition to gravities and non Abelian matter. But here, the dilaton appears only as a substitute
for gravity.

A particular feature of the model to be introduced in the next section is that it features a term that
is higher order in the YM curvature. As will be explained in section {\bf 2}, such terms are necessary
to ensure that the solution yields a finite mass. Such terms were employed in previous
works~\cite{Brihaye:2002hr,Brihaye:2002jg,Breitenlohner:2005hx,Radu:2005mj,Brihaye:2006xc,Radu:2006mb}
with precisely the same purpose. The physical justification for introducing higher order
YM terms, which goes hand in hand with the inclusion of higher order gravitational terms,
is that these occur in the low energy effective action of string theory~\cite{GSW}. Thus
in principle the choice of higher dimensional EYM models involves the selection of
higher order terms in the gravitational and non Abelian curvatures, namely the Riemann and the YM curvatures,
which are reparametrisation and gauge invariant. Because we are concerned with finding classical solutions, we
impose a pragmatic but important further restriction,
namely that we consider only those Lagrange densities that are
constructed from antisymmetrised $2p$ curvature forms, and exclude all other
powers of both Riemann and YM curvature $2$-forms. (In the gravitational
case this results in the familiar Gauss--Bonnet type Lagrangians, while in the case of non Abelian matter to
the YM hierarchy pointed out in footnote $3$ below.) As a result, only
{\it velocity--squared} fields appear in the Lagrangian, which is what is
needed both for physical reasons and for solving the classical field equations.
In practice we add only the minimal number of such higher order terms that
are necessitated by the requirements of finite mass.
This criterion makes the inclusion of higher order gravitational terms
unnecessary since we know from the (numerical) results of \cite{Brihaye:2002hr} that
the qualitative properties of the classical solutions are insensitive to them. In addition to this argument
based on numerical results there is an independent argument advanced at the end of section {\bf 2} of
\cite{Radu:2005mj}, based on the symmetries of the (higher order) gravitational terms, which in the
absence of a dilaton dispenses with the effectiveness of employing such terms. (Note here that the
present YMd model is being used as a prototype for a EYM model, without an additional dilaton field.)
This leaves one with higher order YM curvature terms only, whose status
in the context of the string theory effective action is complex and as yet not fully resolved. While
YM terms up to $F^4$ arise from (the non Abelian version of) the
Born--Infeld action~\cite{Tseytlin}, it appears that this approach
does not yield all the $F^6$ terms~\cite{BRS}.
Terms of order $F^6$ and higher can also be obtained by employing the
constraints of (maximal) supersymmetry~\cite{CNT}. The results
of the various approaches are not identical. In this background, we restrict our considerations to terms in
the YM hierarchy (see footnote $3$) only, in particular to the first two terms. 

Concerning our particular choice of $4+1$ spacetime dimensions here, our reasons are:
When imposing axial symmetry on a YM field in
$d=D+1$ dimensions the simplest way is, following \cite{Witten:1976ck}, to impose
spherical symmetry in the $D-1$ dimensional subspace of the $d$ spacelike
dimensions. In this case the Chern-Pontryagin topological charge is fixed by
the boundary conditions imposed on the first polar angle, and no analogue
of the vortex number appearing in the axially symmetric Ansatz for
$d=3$~\cite{Rebbi:1980yi} is featured~\cite{Witten:1976ck}. Technically, the absence
of a vortex number makes the numerical integration much harder. Imposing
axial symmetry in turns in the $x-y$ and $z-u$ planes of $D=4$ Euclidean
space as in \cite{Radu:2006gg} on the other hand, features two (equal) vortex numbers,
making the numerical work technically more accessible. It is our intention to use the
particular bi-azimuthally symmetric Ansatz of \cite{Radu:2006gg} in $D=4$ that has led us to
restrict ourselves to $d=4+1$ dimensional spacetime. (Numerical work on implementing axial
symmetry like in \cite{Witten:1976ck} is at present in active progress.) Of course, the
exploitation of this type of symmetry is not restricted to $4+1$ spacetime, but
can be extended to any odd $2q+1$ spacetime where $q$ distinct azimuthal symmetries are
imposed, but this in practice results in residual PDE's of order three
and higher for $q\ge 3$.

The second and subsidiary aim of this work is to break the scale invariance
of the usual YM system in $D=4$ studied in \cite{Radu:2006gg}, and the introduction of
the dilaton field does just that. The question of instanton--antiinstanton bound states
in a scale breaking model is an interesing enough matter in itself, presenting a second
important motivation for this work.

In Section 2 we present the model, impose the symmetry and state the boundary
conditions, in successive subsections. The numerical results are presented in
Section 3, presenting both solutions with spherical and bi-azimuthal symmetry.
We give our conclusions and remarks in the final section.

\section{The model}
The model in $5$ spacetime dimensions with coordinates $x_M=(x_0,x_{\mu})$
that we study here is described by the Lagrangian
\be
\label{L}
{\cal L}_m=\frac{1}{4\pi^2}\bigg(|\pa_M\f|^2+
\left(\frac{\tau_1}{2\cdot 2!}\,\e^{2a\f}\,\mbox{Tr}\,
{\cal F}_{MN}^2+
\frac{\tau_2}{2\cdot 4!}\,\e^{6a\f}\,\mbox{Tr}\,{\cal F}_{MNRS}^2\right)\bigg)
\ee
where $\f$ is the dilaton field,
${\cal F}_{MN}=\pa_{M}{\cal A}_{N}-\pa_{N}{\cal A}_{M}+
[{\cal A}_{M},{\cal A}_{N}]$ is the $2$-form YM curvature and
${\cal F}_{MNRS}=\{{\cal F}_{M[N},{\cal F}_{RS]}\}$
is the $4$-form YM curvature consisting of the totally antisymmetrised
product of two YM $2$-form YM curvatures. (The bracket $[\nu\rho\si]$
implies cyclic symmetry.) $\tau_1$ and $\tau_2$ are dimensionful coupling
strengths which will eventually be scaled out against the constant $a$ in
the exponent, which has the inverse dimension of the dilaton field $\f$.
Similar to the $d=3+1$ case, the form we choose for the coupling of the
dilaton field to the nonabelian matter  was found by requiring that
a shift $\phi \to \phi+\phi_0$ of the dilaton field
to be compensated by a suitable rescaling of the coordinates.

 Let us give a brief justification for the choice of the model \re{L}. At the most basic
level it is a YM--dilaton (YMd) model designed to simulate qualitatively a EYM model in $d=5$.
Repacing the dilaton in \re{L} by a gravitational term is a physically relevant model,
representing part of a low energy effective action in $d=5$. Adding gravitational terms to
\re{L} as it stands is a EYMd model, which is just as physically relevant.

The YM system, which scales as $L^{-4}$, in $d=4+1$ supports static solitons, namely the
BPST instantons in $d=4+0$ dimensions. When the usual Einstein--Hilbert gravity, which scales
as $L^{-2}$, is added to the YM term, the soliton collapses because of the (Derrick) scaling
mismatch. To compensate for this scaling mismatch,
a term scaling as $L^{-\nu}$, with $\nu\ge 5$ must be added. If one is to restrict to positive
definite terms~\footnote{The choice of a Chern--Simons term
$\vep_{\la\mu\nu\rho\si}\mbox{Tr}A_{\la}\left(F_{\mu\nu}F_{\rho\si}-F_{\mu\nu}A_{\rho}A_{\si}
+\frac25A_{\mu}A_{\nu}A_{\rho}A_{\si}\right)$ scaling as $L^{-5}$
is a possibility, albeit a considerably harder problem technically, and is at present under
active consideration.}, $\nu$ will be even, and the most economical choice is $\nu=6$. A typical
such term would be $\mbox{Tr}\left(F\wedge DX\right)^2$, where $X$ is a scalar field, {\it e.g.} a
Higgs or sigma--model field. This necessitates the introduction of a completely new (scalar)
field which unlike the dilaton is not directly recognised as a constituent of a low energy
effective action. For this reason we eschew this choice, and restrict our attention instead to
systems featuring only YM (and eventually YMd) fields. The most economical choice then is to
compensate with the YM term $\mbox{Tr}\left(F\wedge F\right)^2$, scaling with $\nu=4$. We note,
finally, that adding a (positive or negative) cosmological constant does not remedy the scaling
mismatch since these terms do not scale at all. Indeed, in all higher dimensional EYM
cases studied, with $\Lambda=0$~\cite{Brihaye:2002hr,Brihaye:2002jg},
$\Lambda<0$~\cite{Radu:2005mj} and $\Lambda>0$~\cite{Brihaye:2006xc},
the mass turns out to be infinite when higher order YM terms are not employed. 

The YM sector of the action density in $4+1$ dimensions employed here, 
is that one used in \cite{Brihaye:2002jg}, namely the superposed $p=1$
and $p=2$ members of the YM hierarchy\footnote{The YM hierarchy labeled by
the integer $p$ was introduced in \cite{Tchrakian:1984gq} in the context of self-dual
solutions in $4p$ Euclidean dimensions, but superpositions of various $p$
members were employed ubiquitously since.}. The dilaton breaks the scale invariance
of the usual $p=1$ YM system, and a simple Derrick-type scaling argument
shows that no finite mass/energy solution can exist if $\tau_2=0$, $i.e.$ the $p=2$  term
in (\ref{L}) is necessary.

The YM and dilaton field equations read
\begin{eqnarray}
\label{YM-eqs}
\tau_1D_{\mu}\left( \e^{2a\f}{\cal F}^{\mu\nu}\right)+
\frac12\tau_2\{{\cal F}_{\rho\si},D_{\mu}\left( e^{6a\f}\
{\cal F}^{\mu\nu\rho\si}\right)\}=0,
\\
\label{dil-eqs}
\nabla^2 \phi= 
\frac{a}{2\pi^2}\left( 2e^{2a\f}\hat L_1+
6e^{6a\f}\hat L_2\right).
\label{YMdeqs}
\end{eqnarray}
In \re{dil-eqs} we have used the notation
\begin{eqnarray}
\hat L_1=\frac{\tau_1}{2\cdot 2!} \mbox{Tr}\,
  {\cal F}_{MN}^2\quad,~~~~\hat L_2=
\frac{\tau_2}{2\cdot 4!}\mbox{Tr}\,{\cal F}_{MNRS}^2\,.\label{hat}
\end{eqnarray}
\subsection{Imposition of symmetry and residual action}
In the YM connection ${\cal A}_{M}=({\cal A}_{0},{\cal A}_{\mu})$,
we choose the temporal component ${\cal A}_{0}=0$ to vanish and
the spacelike components ${\cal A}_{\mu}$ is subjected to
two successive axial symmetries, described in \cite{Radu:2006gg}.
We denote the Euclidean four dimensional coordinates as
$x_{\mu}=(x,y;z,u)\equiv(x_{\al};x_i)$, with $\al=1,2$ and $i=3,4$, 
and use the
following parametrisation
\begin{eqnarray}
x_{\al}=r\sin\ta\,\hat x_{\al}\equiv\rho\,\hat x_{\al},
~~~x_{i}=r\cos\ta\,\hat x_{i}\equiv\si\hat x_i\,,
\label{coord}
\end{eqnarray}
where $r^2=|x_{\mu}|^2=|x_{\al}|^2+|x_{i}|^2$, with the unit vectors
appearing in \re{coord} parametrised as
 $\hat x_{\al}=(\cos\vf_1,\sin\vf_1),~~\hat 
x_{i}=(\cos\vf_2,\sin\vf_2)$, 
with $0\le\ta\le\frac{\pi}{2}$ spanning the
quarter plane, and the two azimuthal angles $0\le\vf_1\le 2\pi$
and $0\le\vf_2\le 2\pi$.
The $d=4+1$ Minkowski spacetime metric reads for this ansatz
\begin{eqnarray}
ds^2=-dt^2+dr^2+r^2(d\theta^2+\sin^2\theta d\vf_1^2+
\cos^2\theta d\vf_2^2).
\end{eqnarray}
The first stage of symmetry imposition is of cylindrical symmetry in the
$x_{\al}=(x_1,x_2)$ plane, and the Ansatz is stated as 
is
\bea
\label{RR-Aa}
&{\cal A}_{\al}=\left(\frac{\phi^5+n_1}{\rho}\right)\,\Sigma_{\al\beta}
\hat x_{\beta}+\left(\frac{\phi^m}{\rho}\right)\,(\vep\hat x)_{\al}\,
(\vep n^{(1)})_{\beta}\,\Sigma_{\beta m}
+A_{\rho}^{m5}\,\,\hat x_{\al}n^{(1)}_{\beta}\,\Sigma_{\beta m}
-A_{\rho}^{34}\,\hat x_{\al}\Sigma_{34}~,
\\
&{\cal A}_i=A_i^{m5}\, 
n^{(1)}_{\beta}\,\Sigma_{\beta m}-A_i^{34}\,\Sigma_{34}~,
\label{RR-Ai}
\eea
in which the index $m=3,4$ is summed over, and the unit vector
$n^{(1)}_{\al}=(\cos n_1\vf_1\,,\,\sin n_1\vf_1)$ is
labeled by the vortex integer $n_1$, $\vep_{\al\beta}$ being the
Levi-Civita symbol. The spin matrices
$\Sigma_{\mu\nu}=(\Sigma_{\al\beta},\Sigma_{\al i},\Sigma_{ij})$ in
\re{RR-Aa} and \re{RR-Ai} are one or other of the two
chiral representations of $SO(4)$, $i.e.$ they are $SU(2)$ matrices.

If in \re{RR-Aa}-\re{RR-Ai} we regard the functions $(\f^m,\f^5)\equiv\f^a$
as a $SO(3)$ isovector field, and $(A_{\rho}^{m5},A_{\rho}^{34})\equiv A_{\rho}^{ab}$ and
$(A_{i}^{m5},A_{\rho}^{34})\equiv A_{i}^{ab}$ as $SO(3)$ YM connection fields, with
antisymmetric $SO(3)$ algebra indices $[ab]=([34],[45],[53])$, then it turns out that
${\cal F}_{\mu\nu}=({\cal F}_{\al\beta},{\cal F}_{\al i},{\cal F}_{ij})$
is expressed exclusively in terms of the curvature
$(F_{ij}^{ab},F_{i\rho}^{ab})$ of the $SO(3)$ connection
$(A_{\rho}^{ab},A_{i}^{ab})$, and the corresponding covariant derivative
$(D_i\f^a,D_{\rho}\f^a)$ of $\f^a$, all defined on the hyperbolic space
with coordinates $(x_i,x_{\rho})$.

The second stage of symmetry imposition is expressed most succintly by rewriting the
residual fields $(A_{i}^{ab},A_{\rho}^{ab})$ and $\f^a$ on this hyperbolic space in matrix
representation
\bea
A_i=-\frac12\,A_{i}^{ab}\,\Sigma_{ab}\quad&,&\quad
A_{\rho}=-\frac12\,A_{\rho}^{ab}\,\Sigma_{ab}~,\label{matrixa}\\
\Phi&=&\f^a\,\Sigma_{a4}\,.\label{matrixf}
\eea
Azimutal symmetry in the $(x_3-x_4)$ plane is imposed on the connection fields
\re{matrixa} by decomposing $A_i$ formally according to \re{RR-Aa} and
$A_{\rho}$ according to \re{RR-Ai}. Noticing now that the index $a$ in
\re{matrixa}-\re{matrixf} runs only over three values, and reassigning the
values of the index $i=1,2$, the analogues of \re{RR-Aa} and \re{RR-Ai}
contract to give
\bea
\label{RR2-Aa}
A_{i}&=&\left(\frac{\chi^4+n_2}{\si}\right)\,\Sigma_{ij}
\hat x_{j}+\left(\frac{\chi^3}{\rho}\right)\,(\vep\hat x)_{i}\,
(\vep n^{(2)})_{j}\,\Sigma_{j3}
+A_{\si}^{34}\,\,\hat x_{i}n^{(2)}_{j}\,\Sigma_{j3}~,
\\
A_{\rho}&=&A_{\si}^{34}\, 
n^{(2)}_{j}\,\Sigma_{j m}~,
\label{RR2-Ai}
\eea
exhibiting the Abelian connection $A_{\si}^{34}$ analogous to the
Abelian connection $A_{\rho}^{34}$ appearing in \re{RR-Aa} and the isodoublet
function $(\chi^3,\chi^4)$.

The corresponding axially symmetric decomposition of $\Phi$ in \re{matrixf} is
\be
\label{axf}
\Phi=\xi^{1}\,n^{(2)}_j\,\Sigma_{j4}+\xi^{2}\,\Sigma_{34}\,.
\ee
In \re{RR2-Aa}-\re{RR2-Ai} and \re{axf} we have used the unit vector
$n^{(2)}_{i}=(\cos n_2\vf_2\,,\,\sin n_2\vf_2)$, with
vorticity integer $n_2$. The final stage of symmetry imposition is to treat
the two azimuthal symmetries imposed in the $x-y$ and the
$z-u$ planes on the same footing, leading to the equality of the two
vortex numbers, $n_1=n_2\equiv n$.

Denoting the residual functions
$(A_{\rho}^{34},A_{\si}^{34})=(a_{\rho}a_{\si},)$,
$(\chi^3,\chi^4)=\chi^A$, $(\xi^1,\xi^2)=\xi^A$, and regarding
$(a_{\rho},a_{\si})$ as an Abelian connection on the quater plane defined
by $(\rho,\si)$, the residual action densities can be expressed exclusively
in terms of the $SO(2)$ curvature
\be
\label{finalf}
f_{\rho\si}=\pa_{\rho}a_{\si}-\pa_{\si}a_{\rho}
\ee
and the covariant derivatives
\begin{eqnarray}
{\cal D}_{\rho}\chi^A=\pa_{\rho}\chi^A+
a_{\rho}(\vep\chi)^A\quad,\quad
{\cal D}_{\si}\chi^A=\pa_{\si}\chi^A+
a_{\si}(\vep\chi)^A,
\label{dchi}
\\
{\cal D}_{\rho}\xi^A=\pa_{\rho}\xi^A+
a_{\rho}(\vep\xi)^A\quad,\quad
{\cal D}_{\si}\xi^A=\pa_{\si}\xi^A+
a_{\si}(\vep\xi)^A.
\nonumber
\end{eqnarray}
The residual two dimensional YM action densities descending from the $p=1$
and the $p=2$ terms $\hat L_1$ and $\hat L_2$ defined by \re{hat} are, respectively,
\bea
&&L_1 = \frac{\tau_1}{4}\left[\rho\si\,f_{\rho\si}^2
+\frac{\rho}{\si}\left(|{\cal D}_{\rho}\chi^A|^2
+|{\cal D}_{\si}\chi^A|^2\right)
+\frac{\si}{\rho}\left(|{\cal D}_{\rho}\xi^A|^2
+|{\cal D}_{\si}\xi^A|^2\right)+
\frac{1}{\rho\si}(\vep^{AB}\chi^A\xi^B)^2\right],
\label{redact1}
\nonumber
\\
&&L_2 =
\frac{\tau_2}{12\,\rho\si}\,\left(\vep_{AB}\chi^A\xi^B\,f_{\rho\si}+
{\cal D}_{[\rho}\chi^A\,{\cal D}_{\si]}\xi^A\right)^2.
\label{redact12}
\eea
These residual action densities are scalars with
respect to the {\it local} $SO(2)$ indices $A,B$, hence they are
manifestly gauge invariant. They describes a $U(1)$ Higgs like model 
with {\bf two} effective Higgs fields $\chi^A$ and $\xi^A$, coupled minimally
to the $U(1)$ gauge connection $(a_{\rho},a_{\si})$. To remove this
$U(1)$ gauge freedom we impose the usual gauge condition
\be
\label{gc}
\pa_{\rho}a_{\rho}+\pa_{\si}a_{\si}=0~.
\ee
Since our numerical constructions will be carried out using the 
coordinates $(r,\ta)$  we display  \re{redact12} also as
\bea
L_1&=&\frac{\tau_1}{4}\bigg[  r\sin\ta\cos\ta\,f_{r\ta}^2
+\frac{r\sin\ta}{\cos\ta}\left(|{\cal D}_{r}\chi^A|^2
+\frac{1}{r^2}|{\cal D}_{\ta}\chi^A|^2\right)\nonumber
\\
&&\qquad\qquad+\frac{r\cos\ta}{\sin\ta}\left(|{\cal D}_{r}\xi^A|^2
+\frac{1}{r^2}|{\cal D}_{\ta}\xi^A|^2\right)+
\frac{1}{r\sin\ta\cos\ta}(\vep^{AB}\chi^A\xi^B)^2\bigg]\label{redactsph1}\\
L_2&=&\frac{\tau_2}{12\,r^3\,\sin\ta\cos\ta}\left(\vep_{AB}\chi^A\xi^Bf_{r\ta}
+{\cal D}_{[r}\chi^A{\cal D}_{\ta]}\xi^A\right)^2~.
\label{redactsph2}
\eea
The total mass-energy $M$ of the system is
\begin{eqnarray}
M=\int d^4x\sqrt{g}\ \mathcal{L}_m=
 \int_0^{\infty}dr\int_0^{\pi/2}d
\theta~\left[\frac{1}{2}r^3 \sin \theta \cos \theta (\phi_{,r}^2+
\frac{1}{r^2}\phi_{,\theta}^2)+ (e^{2a\phi}L_1+e^{6a\phi}L_2)\right],
\end{eqnarray}
and equals the total action of solutions, viewed as solitons 
in a $d=4$ Euclidean space.
  
\subsection{Boundary conditions}
 To obtain regular solutions with finite energy density we impose at 
the origin 
($r=0$) the boundary conditions  
\be
\label{r0}
a_r=0\quad,\quad a_{\ta}=0\quad,\quad
\chi^A=\left(
\begin{array}{c}
\ 0 \\
-n_2
\end{array}
\right)\quad,\quad
\xi^A=\left(
\begin{array}{c}
\ 0 \\
-n_1
\end{array}
\right)\,~,
\ee
which are requested by the analyticity of the YM ansatz,
 and $\partial_r\phi|_{r=0}=0$
for the dilaton field.
In order to find finite mass solutions, we impose at infinity 
\be
\label{rinfty}
a_r=0,~a_{\ta}=-2m,
~\chi^A=
(-1)^{m+1}n_2\,\left(
\begin{array}{c}
\sin 2m\ta \\
\cos 2m\ta
\end{array}
\right),~
\xi^A=-n_1\,\left(
\begin{array}{c}
\sin 2m\ta \\
\cos 2m\ta
\end{array}
\right)\,,~\phi=0,
\ee
$m$ being a positive integer. Similar considerations lead to the following
boundary conditions on the $\rho$ and $\sigma$ axes:   
\bea
a_r=\frac{1}{n_1}
\pa_{r}\xi^1,~~
a_{\ta}=\frac{1}{n_1}
\pa_{\ta}\xi^1,~~
\chi^1=0,~~
\xi^1=0,~~
\pa_{\ta}\chi^2=0,~~
\xi^2=-n_1,
~\partial_\theta \phi=0,
\label{th0}
\eea
for $\theta=0$, and  
\bea
\label{thp2}
a_r=\frac{1}{n_2}
\pa_{r}\chi^1,~~
a_{\ta}=\frac{1}{n_2}
\pa_{\ta}\chi^1,~~
\chi^1=0,~~
\xi^1=0,~~
\chi^2=-n_2,~~
\pa_{\ta}\xi^2=0,~~
\partial_\theta \phi=0,
\eea
for $\theta=\pi/2$, respectively.
\subsection{Topological charge}
 In our normalisation, the topological charge is defined as
\be
\label{pont}
q=\frac{1}{32\pi^2}\vep_{\mu\nu\rho\si}\int\mbox{Tr}\{
{\cal F}_{\mu\nu}{\cal F}_{\rho\si}\}\,d^4x\,,
\ee
which after integration of the azimuthal angles $(\vf_1,\vf_2)$ 
reduces to
\bea
q&=&\frac12\vep_{\mu\nu}\,\int\left(\frac12\,\vep_{AB}\chi^A\xi^B\,f_{\mu\nu}+
{\cal D}_{\mu}\chi^A\,{\cal 
D}_{\nu}\xi^A\right)\,d^2x\label{topch2}\\
&=&\frac14\int\vep_{\mu\nu}\,\pa_{\mu}(\chi^A{\cal D}_{\nu}\xi^A-
\xi^A{\cal D}_{\nu}\chi^A)\,d^2x\,.\label{totdiv}
\eea
The integration in \re{topch2} is carried out over the 2 dimensional
space $x_{\mu}=(x_{\rho},x_{\si})$. 
As expected this is a total
divergence expressed by \re{totdiv}.
 
Using Stokes' theorem, the two dimensional integral of \re{totdiv} 
reduces
to the one dimensional line integral
\bea
\label{stokes}
q=\frac14\int
\chi^A\stackrel{\leftrightarrow}{\cal D}_{\mu}\xi^A\,ds_{\mu},
\eea
This integral has been evaluated in \cite{Radu:2006gg}
 by reading off the appropriate values 
of
$\chi^A$ and $\xi^A$ from \re{th0}-\re{thp2}. The result is
\be
\label{q}
q=\frac12\,[1-(-1)^m]n_1n_2.
\ee
 
\section{Numerical results}
Apart from the coupling constants $\tau_1$ and $\tau_2$ the model contains also
the dilaton constant $a$.
Dimensionless quantities are obtained by rescaling 
\begin{eqnarray} \label{tau}
\phi \to \phi/a,~~r \to r (\tau_2/\tau_1)^{1/4}, 
\end{eqnarray} 
This reveals the existence of 
one fundamental parameter which gives the strength of
the dilaton-nonabelian interaction
\begin{eqnarray}
\alpha^2=a^2 \tau_1^{3/2}/\tau_2^{1/2}\,,\label{alpha}
\end{eqnarray}
which is a feature present also in the  EYM case \cite{Brihaye:2002jg}.
We use this rescaling to set $\tau_1=1$, $\tau_2=1/3$ in the numerical
computation, without any loss of generality.
 
One can see that the limit $\alpha \to 0$ can be approached in two ways
and two different branches of solutions may exist. The first limit corresponds to a pure $p=1$
YM theory with vanishing dilaton and $p=2$ YM terms, the solutions here replicating the
(multi-)instantons and composite instanton-antiinstanton bound states
discussed in \cite{Radu:2006gg}. The other possibility
corresponds to a finite value of the dilaton coupling $a$ as $\tau_1 \to 0$. Thus, the 
second limiting configuration is a solution of the truncated $p=2$ YM system interacting
with the dilaton, with no $p=1$ YM term.

We have studied YMd solutions with $m=1,~2$. From our knowledge of the
tolopogical charges \re{q}, the $m=1$ solutions will describe 
(multi)solitons and the $m=2$ solutions, soliton-antisoliton
configurations. Also, to simplify the general picture we set $n_1=n_2=n$
in the boundary conditions (\ref{r0})-(\ref{thp2}). 

The spherically symmetric solutions are found
by using a standard differential equations solver. 
The numerical calculations in the bi-azimuthally symmetric case
were performed with the software 
package CADSOL, based on the Newton-Raphson method \cite{FIDISOL}. 
In this case, the field equations are first discretized on a 
nonequidistant grid and the resulting system
is solved iteratively until convergence is achieved.
In this scheme, a new radial variable $x=r/(1+r)$ is introduced
which maps the semi-infinite region $[0,\infty)$ to the closed region $[0,1]$.

As will be described below, solutions exist for certain ranges of the parameter $\al$. It
turns out that $m=1$ solutions with all $n$ and $m=2$ solutions with $n=1$ exist for a
range of $\al$ starting from a $\al\to 0$ limit, but do not persist
all the way up to the second $\al\to 0$ limit.
(However, the way the solutions approach the limit $\al\to 0$ depends on $m$.)
 By contrast we find that
$m=2$ solutions with all $n>1$, exist for all $\al$ between the two limits.

\subsection{$m=1$ configurations}

 \subsubsection*{$n=1$ spherically symmetric solutions}

In the spherically symmetric limit, which case we shall analyse numerically
first, the angular dependence of these functions is fixed and the only
remaining independent function depends on the variable $r$.
The independent function
in this case is $a_{\ta}=w(r)-1$, with the remaining fuctions
$(a_r,\chi^A,\xi^A)$ given by
\be
\label{m1n1}
a_r=0,~\chi^1=-\xi^1=\frac{1}{2}(w(r)-1) \sin 2 
\theta,~~
\chi^2=-(w(r)-1) \cos^2 \theta-1,~~\xi^2=-(w(r)-1) \sin^2 \theta-1.
\ee  
The functions $\phi(r)$ and $w(r)$ satisfy the equations
\begin{eqnarray}
 (r^3\phi')'=\alpha^2 \left(2e^{2\phi}\tau_1 (rw'^2+\frac{(w^2-1)^2}{r})
+9e^{6 \phi}\tau_2\frac{(w^2-1)^2}{r^3}w'^2\right),\nonumber
\\
\left(e^{2\phi}rw'(\tau_1+3e^{4\phi}\tau_2\frac{(w^2-1)^2}{r^4})\right)'=
\frac{2e^{2\phi}w(w^2-1)}{r}\left(\tau_1+
3\tau_2e^{4\phi}\frac{w'^2}{r^2}\right)\label{eqs-sph}.
\end{eqnarray}
The asymptotic solutions to these functions can be systematically
constructed in both regions, near the origin 
and for $ r \gg 1$.
The small $r$ expansion is
\begin{eqnarray}
\label{sig0}
w(r)=1-br^2+O(r^4),~~
\phi =\phi_0+4\alpha^2(\frac{\tau_1}{2}+9\tau_2b^2)b^2 r^2+O(r^4),
\end{eqnarray}
with $b$, $\phi_0$ two real parameters,
while as $r \to \infty$ we find
\begin{eqnarray}
\label{sig1}
w(r)=\pm 1-\frac{4\phi_1}{r^2}-\frac{4\phi^3_1}{27r^6}+O(1/r^8),~~
\phi = 
\frac{ \phi_1}{r^2}-\frac{32\alpha^2\phi^3e^{2\phi_0}\tau_1}{27r^6}+O(1/r^8).
\end{eqnarray}
We numerically integrate the Eqs. \re{eqs-sph} 
with the above set of boundary conditions
 for 
$\tau_1=1,\tau_2=1/3$ and varying $\alpha$.
The picture we found is very similar to that found for the EYM
system \cite{Brihaye:2002jg},
the dilaton coupling constant playing the role of the Newton constant.
First, for a given $\alpha$, 
solutions with the right asymptotics
exist for a single value of the "shooting" parameter
$b$ which enters the expansion $(\ref{sig0})$.
For $\alpha$ small enough, a branch of
solutions smoothly emerges from the BPST configuration \cite{BPST}.
When $\alpha$ increases,  the mass  $M$ and the absolute value of the
dilaton function at the origin increase, 
as indicated in Figure~1. These solutions exist up to a maximal
value $\alpha_{max}\simeq 0.36928$ of the parameter $\alpha$.

As in the corresponding gravitating case~\cite{Brihaye:2002jg}, we found another branch of
solutions in the interval $\alpha \in [\alpha_{cr(1)} , \alpha_{max}]$
with $\alpha_{cr(1)}^2\simeq 0.2653$.
On this second branch of solutions, both $\phi(0)$ and $M$ continue
to increase but stay finite. However, a third branch of solutions exists
for $\alpha \in [0.2653, 0.2652]$, on which the two quantities increase further.
A fourth branch of solutions has also been
found, with a corresponding $\alpha_{cr(3)}\simeq 0.2642$. 
The mass $M$, the value of the dilaton field at the origin
$\phi(0)$ and the initial (shooting) parameter $b$ increase along these branches.
Further branches of solutions, exhibiting more
oscillations around $\alpha \simeq 0.264 $ are very likely
to exist but their study is a difficult numerical problem.
This critical behaviour is described
as a {\it conical fixed point} in the analytic analysis in \cite{Breitenlohner:2005hx}.
Therefore we conclude that, as in the spherically symmetric gravitating
case~\cite{Brihaye:2002jg}, the limit $\tau_1=0$ is not approached for solutions with $m=1,~n=1$.

As a general feature, all solutions discussed here present only one 
node in the gauge function $w(r)$. As in the higher dimensional EYM models discussed in
\cite{Brihaye:2002hr,Breitenlohner:2005hx}, no multinode solutions were found.

\subsubsection*{$n>1$ }

Solutions with bi-azimuthal symmetry with nontrivial
dependence on both $r$ and $\theta$ are found for $(n_1,n_2) \neq 1$
subject to the boundary conditions (\ref{r0})-(\ref{thp2}). We have studied 
solutions for $m=1$ with  $2\leq n\leq 5$.
The general features of the $m=1$ solutions are the same for all $n>1$.
Also, as seen in (\ref{q}), the $m=1$ configurations carry a topological 
charge $q=n^2$. 
The corresponding solutions of the ${\cal F}_{MN}^2$ model are self-dual
and have been considered already in \cite{Brihaye:1982cc}, \cite{Brihaye:1989uw}
(for a different parametrization of the gauge field, however).

These solutions are constructed by starting with the known spherically 
symmetric configuration and increasing the winding number $n$ in small steps.
The iterations converge, and repeating the procedure one obtains
in this way solutions for arbitrary $n$. The physical values of $n$ are integers.
The typical numerical error for the functions is estimated to be 
of the order of $10^{-3}$ or lower.

Any spherically symmetric configuration
appears to result in  generalisations with higher winding numbers $n$. 
Moreover, the branch structure noticed for the $m=1,~n=1$ case
seems to be retained by the higher winding number $m=1$ solutions.
Again, the first branch of solutions exists up to a maximal value of $\alpha$,
where another branch emerges, extending backwards in $\alpha$.
We managed to construct higher winding number $n$ counterparts
of the first two branches of spherically symmetric solutions.
The mass  $M$ and the absolute value of the
dilaton function at the origin increase along these branches, as shown in Figure 2.
Note that the value of the dilaton function at the origin
exhibited in the figures is actually $\phi(r=0,\ta=0)$, restricting to $\ta=0$.
This restriction is reasonable since for all solutions with bi-azimuthal symmetry discussed in
this paper, the dilaton function at $r=0$ presents almost no dependence on the angle $\theta$.

We expect that the oscillatory pattern of $\f(0)$ arising from the {\it conical fixed point}
observed above for the spherically symmetric $n=1$ solutions, will also be discovered for the
$n>1$ solutions here, but their construction is a
difficult numerical problem beyond the scope of the present work.
 
In Figure 3 we present the gauge functions, the dilaton,
and the topological charge density
\[
\varrho=\frac14\vep_{\mu\nu}\,\left(\vep_{AB}\chi^A\xi^B\,f_{\mu\nu}+
{\cal D}_{[\mu}\chi^A\,{\cal D}_{\nu]}\xi^A\right)
\]
read off (\ref{topch2}), as functions of the radial coordinate
$r$ for five different angles for a typical first branch $m=1,~n=3$ solution
with $\alpha=0.21$. The functions $a_\theta$ and $\phi$ have a small $\theta$ dependence  
(although the angular dependence increases with $n$), 
while $\chi_1$ and $\xi_1$ have rather similar shapes.
The action density $\mathcal{L}$ possesses one maximum on the $\theta=\pi/4$ axis. All
multicharge solutions found have concentrated energy and charge density profiles where
individual (unit) charge consituents do not appear as distinct components.
The moduli of the effective Higgs fields $|\chi|=(\chi^A\chi^A)^{1/2}$ and
$|\xi|=(\xi^A\xi^A)^{1/2}$  possess one node each on the $\rho$ and the $\sigma$ axes,
respectively, which coincide with the maximum of the action density. 
The position of this node moves inward along the first and second branches.

 
\subsection{$m=2$ configurations}
The $m=2$ configurations can be thought of as composite systems consisting of two components which
are pseudoparticles of topological charges $\pm n$. Thus, these configurations reside in the
topologically trivial sector and carry no  Chern-Pontryagin topological charge. This type of
solutions have no spherically symmetric limit. Also, their behaviour as a function of $\alpha$ is
different from those with $m=1$ presented above, in that solutions for all values of $\al$
exist between the two distinct limits of $\al\to 0$ implied by \re{alpha},
for all $n$ except for $n=1$.

\subsubsection*{$n=1$}
It is perhaps interesting to note from the outset that $m=2\ ,\ n=1$ solutions to be described
now, have apparently no counterpart in the $4+0$
dimensional $p=1$ YM model studied in \cite{Radu:2006gg}. (It turns out that for
the $(m=2,n=1)$ solution in that case there is no analytic proof of existence either.)
The obvious difference of the $4+1$ dimensional model \re{L} here and the $4+0$
dimensional $p=1$ YM model is that the solutions of the former are parametrised by
the effective coupling constant $\alpha$, while the latter has no such parameter. As will be
described below, $m=2\, \ n=1$ solutions exist for a certain range of $\al$, and this range
excludes the limiting case where the contribution to the action of
the dilaton term and the $p=2$ YM term in \re{L}  disappear,  $i.e.$ 
a ${\cal F}_{MN}^2$ model. 

We find that in the limit $\al\to 0$ resulting from $a\to 0$, {\it cf.} \re{alpha},
no solutions of this type exist. However in the limit $\alpha \to 0$ corresponding to a finite
value of the dilaton coupling $a$ as $\tau_1 \to 0$, such solutions exist.
This limiting configuration is then a solution of the truncated system consisting of the dilaton
term and $p=2$ YM term ${\cal F}_{MNRS}^2$, which dominate. Its characteristic 
feature is that for this configuration both nodes of the effective Higgs 
fields $|\chi| $ and $|\xi|$  merge on the $\theta = \pi/4$ axis. 
A family of solutions of the model (\ref{L}) emerges from this configuration.
As $\alpha$ increases, the nodes move towards the symmetry axes, $\rho$ and $\sigma$, 
respectively, forming two identical vortex rings whose radii slowly decrease while
the separation of both rings from the origin increase. At the critical value $\alpha_{cr}\simeq 0.265$
the node structure of the configuration changes, both vortex rings shrink to zero size and two 
isolated nodes appear on each symmetry axis. This structure is known for the usual YM system 
in $d=4+0$ \cite{Radu:2006gg}, indeed, increasing of $\alpha$ along this branch can be associated
with increasing of the coupling $\tau_1$ w.r.t. $\tau_2$ as the dilaton coupling $a$ remains
fixed; then the term ${\cal F}_{MN}^2$ becomes leading. The maximum of the action density however
is still located on  $\theta = \pi/4$ axis.

Another similarity with the instanton-antiinstanton solution of the $d=4+0$ $p=1$ YM theory is
that the gauge functions  $a_r$, $a_\theta$ as well the dilaton function $\phi$ of the $n=1, m=2$
solutions also are almost $\theta$-independent. Along this branch the mass of the solutions grows 
with increasing $\alpha$ since with increasing coupling  $\tau_1$ the contribution of the 
term ${\cal F}_{MN}^2$ also increases.

As the effective coupling increases further beyond  $\alpha_{cr}$ the relative distance between 
the nodes increases, one lump moving towards the origin while the other one moves in the opposite 
direction. Along this branch  both the value of the dilaton field at the origin $|\phi(0)|$ and 
mass of configuration $M$ increase as $\alpha$ increases. This branch 
extends up to a maximal value $\alpha_{max}^{(1)}\simeq0.311$ beyond which 
the dilaton coupling becomes too strong for the static configuration to persist. The second
branch, whose energy is higher, extends backwards up to  $\alpha_{max}^{(2)}\simeq 0.279$.
Along this branch both $|\phi(0)|$ and
the mass of the configuration continue to increase as $\alpha$ decreases. 
 Also the separation between the nodes decreases and
both nodes invert direction of the motion, moving 
toward
each other along this branch.  In Figure 5 we present 
the values of the dilaton function  at the origin $\phi(0)$ and the total
mass (rescaled by $\al^2$) of these configurations as functions of $\alpha$.

\subsubsection*{$n=2$}

This configuration also resides in the topologically trivial sector and can be considered as 
consisting of two solitons of charges $n=\pm 2$. Then the interaction between the nonabelian 
matter fields becomes stronger than in the case of unit charge constituents and the expected
pattern of possible branches of solutions is different from the $n=1$ case above.

Indeed, the $n=2\ , m=2$ solutions show a different dependence on the coupling constant $\alpha$,
with two branches of solutions. The lower branch emerges from the
corresponding solution in pure $p=1$ YM theory with vanishing dilaton and $p=2$ YM terms,
replicating the corresponding solution in \cite{Radu:2006gg}. The variation of the 
effective coupling along this branch is associated with the decrease of
$\tau_1$, at fixed $\tau_2$ and fixed dilaton coupling $a$.
The second branch emerges from a solution of the $p=2$ YM-dilaton system, 
the unrescaled mass $M$ diverging in this limit, with the rescaled mass $M\alpha^2$ vanishing as
seen from Figure 5a. At the maximal value $\alpha_{max} \simeq 0.2372$ this branch bifurcates with
the lower YM branch. For larger values of $\alpha$,
the dilaton coupling becomes too strong for the static configurations to persist.
Thus for $0\leq \alpha < \alpha_{max}$ we notice the existence of 
(at least) two distinct solutions for the same value of coupling constant.

For the same value of $\alpha$, the mass of the  second branch solution
is larger that that of the corresponding lower branch configuration(s).
One should also notice the existence of a curious backbending of the
lower branch for $0.193<\alpha<0.218$. Four distinct solutions exist in this case for the
same value of $\alpha$ (three of them located on the lower branch),
distinguished by the value of the mass and the dilaton field at the origin.
This pattern is illustrated in Figure 5.

Again, observation of the positions and structure  
of the nodes of the effective scalar fields allows 
us to better understand the behaviour of the 
solutions. For lower branch solutions with
small values of $\alpha$ there are two (double) nodes of the 
fields  $|\chi| $ and $|\xi|$ on the $\rho$ and $\sigma$ symmetry axes respectively.  
The locations of nodes correspond to the locations of the two individual 
constituents and the action density distribution posesses two distinct maxima on 
the $\theta = \pi/4$ axis. The  distance
between these nodes changes only slightly along the lower mass branch.
The backbending in $\alpha$ observed in this case is reflected also for in the
relative positions of the nodes. At the maximal value of $\alpha$, the inner node is located at 
$\rho_0^{(1)} = \sigma_0^{(1)} \simeq 2.97$ 
and the outer node is located at $\rho_0^{(2)} = \sigma_0^{(2)} \simeq 4.18$.
 
Along the upper branch, as $\alpha$ slightly decreases below $\alpha_{max}$, 
the inner node inverts direction of its movement toward the outer node which still moves inwards. Thus, 
both nodes on the symmetry axis rapidly approach each other and merge forming a two vortex ring
solution as $\alpha \simeq 0.2355$. 
The action density then has a single maximum on $\theta = \pi/4$ axis. As $\al$ decreases further
both nodes move away from the symmetry axis and their positions do not coincide 
with the location of the maximum of the action density. 
Further decreasing $\alpha$  results in the increase of the radii of the two rings around the symmetry
axis, and in the limit  $\alpha\to0$ the rings touch each other on the $\theta = \pi/4$ axis. 
 
In Figure 4 we give three dimensional plots of
the modulus of the effective Higgs field $\xi$ for the $n=m=2$ upper branch vortex solution 
at $\alpha = 0.20$ and the $n=m=2$ lower branch double node solution at the same value of 
$\alpha$. The action density as given by (\ref{L}) is also plotted at $\alpha = 0.20$ both 
for the upper and for the lower branches.

The numerical calculations indicate the possibility that the solutions of the
fundamental YM branch, namely the branch on which the $p=1$ YM term dominates, are not unique.
It is possible that higher linking number 
configurations with higher masses might exist. This possibility will be explored elsewhere. 

\subsubsection*{$n=3$}

For the $n=3$ configuration composed of two triple charged pseudoparticles we 
observe a somewhat simpler pattern. The lower dilaton branch emerges from the 
limit of vanishing dilaton and $p=2$ YM couplings and extends up to a maximal value
$\alpha_{cr} = 0.165$. Along this branch the configuration posesses two vortex rings. 

As $\alpha$ increases the mass of the solution increases and, at the same time, the radii of the
rings slowly increase and 
both rings move inwards. This lower mass branch bifurcates at the critical value of the effective 
coupling $\alpha_{cr}$ with an upper branch which extends all the way back to $\alpha = 0$
(see Figure 5).  Again, in this limit both vortex rings come into contact on the $\theta=\pi/4$ axis.
Thus, we observe no isolated nodes on the symmetry axis and both upper and lower energy branches
correspond to vortex ring solutions.

Note that the branch structure here closely resembles the pattern which was observed 
for the gravitating monopole-antimonopole chains and vortex solutions in $d=3+1$ 
Einstein-Yang-Mills-Higgs system~\cite{Kleihaus:2003nj, Kleihaus:2004fh}. 

The profiles of a typical $m=2,~n=3$ solution are presented in Figure 6
(the picture there applies as well for  $n=1,~2$ configurations).

\section{Summary and discussion}

Finite mass static solutions to a $4+1$ dimensional $SU(2)$ YMd model are
constructed numerically. The YM sector of the model consists of
the usual YM term, labeled $p=1$, and the next higher order $p=2$ term of the YM
hierarchy~\cite{Tchrakian:1984gq}. The second YM term is necessary to counteract the
scaling of the   quadratic  dilaton   kinetic 
 term, since we require finite energy solutions. 
 In section {\bf 2} we have explained the rationale behind our choice of the model \re{L}.
Basically our criterion is that of compensating for the scaling of the quadratic kinetic dilaton term, by
introducing an additional term that scales as $L^{-\nu}\ ,\ \  (\nu\ge 5)$, with the further criteria that
this additional term be positive definite, and, that no new fields beyond the dilaton and the YM be
employed.

The solutions constructed are subject to the bi-azimuthal symmetry applied in \cite{Radu:2006gg}.

Viewed as a $4+0$ dimensional model, this is a scale breaking version of a $p=1$ YM theory whose
bi-azimuthally symmetric solutions were presented in \cite{Radu:2006gg}. The latter model being
scale invariant in $d=4+0$, the present solutions have the new feature that they are referred to an
absolute scale. This is an interesting feature of the present work.

Our main motivation here, however, is to study a system which can give an insight into the qualitative
features of static, finite mass solutions to a gravitating YM system in higher dimensions, which
are not subject to spherical symmetry. To date the only higher dimensional Einstein--Yang-Mills (EYM) solutions
known~\cite{Brihaye:2002hr,Brihaye:2002jg,Breitenlohner:2005hx,Radu:2005mj,Brihaye:2006xc,Radu:2006mb} are
subject to spherical symmetry in the spacelike dimensions. Rather than tackle the appreciably more
complex numerical problem of constructing non spherically symmetric EYM solutions, we consider
here instead the corresponding YM--dilaton problem, knowing that the classical solutions of
the latter simulate~\cite{Maison:2004hb} the qualitative properties of the corresponding EYM ones
in $d=3+1$ dimensional spacetime. Thus the present work is a warmup to the ultimate aim of
constructing non-spherically symmetric 
EYM solutions in higher dimensions. Having said that, we
note that we have already made an appreciable start in the construction of the corresponding
$d=4+1$ EYM solutions, and our results to date confirm the qualitative 
similarity of those with the dilaton--YM
solutions presented here. Our results on the gravitating system will be reported elsewhere.

In the context of giving a qualitative description of our results, it must be noted that in this first
effort, we have restricted the dimensionality of the spacetime to $4+1$. Thus we can only compare our
results here with the $4+1$ dimensional subset of the spherically symmetric EYM
solutions~\cite{Brihaye:2002hr,Brihaye:2002jg,Breitenlohner:2005hx,Radu:2005mj,Radu:2006mb},
which were given in spacetime dimensions $d\ge 5$.
We would expect however that the comparative features between spherically
and the non spherically symmetric solutions which we uncover here, 
will stay qualitatively valid also in dimensions $d\ge 5$.

As it happens, spherically symmetric EYM solutions in $4+1$ dimensions~\cite{Brihaye:2002jg} exhibit
quite different qualitative properties compared to those in $5+1$, $6+1$ and $7+1$
dimensions~\cite{Brihaye:2002hr}. Indeed, the comparative patterns remain true modulo $4p$
dimensions, as explained in \cite{Breitenlohner:2005hx}. These features in question concern the
branch structure of the said solutions with respect to the effective coupling parameter $\al$ in the
problem. It turns out that for an appropriate EYM model in $d=4p+1$ dimensions, a peculiar branch
structure, absent in $d\neq 4p+1$, occurs. This was explained in \cite{Breitenlohner:2005hx} to be due
to the occurence of what was called there a {\it conical fixed point singularity}, which manifests
itself by the oscillatory behaviour in $\alpha$ of the global quantities near the critical value of
 the effective coupling parameter\footnote{
This feature persists also when a scalar matter field is added, for
example when a gauged Grassmannian sigma model field is included in the
Lagrangian~\cite{Brihaye:2004gv}. We expect it to persists also when a Higgs field is added instead,
in $4p+1$ dimensions.}.

Now in the present work we do not employ gravity but have instead the
dilaton, which is represented by the singlet scalar field $\f$. It is therefore unavoidable that the
role of the metric function at the origin in the gravitating case, should be replaced here by
$\f(r=0)$, the dilaton function at the origin. This correspondence can be made uniquely in the special
case of our $(m=1\ ,\ n=1)$ bi-azimutal solutions, which are simply the spherically symmetric subset.
As seen from Figure 1, the oscillatory pattern of $\f(0)$ is clear.
 
Concerning $(m\ge 2\ ,\ n)$ solutions, these are not spherically symmetric and depend on $\ta$ in
addition to $r$. Instead we have employed the values $\f(r=0,\ta=0)$ to track the branch behaviour of
our solutions. This is quite a reasonable criterion, since we find that the dependence of $\f(0,\ta)$
is very small on all branches constructed. The branch structures for $(m=2\ ,\ n=1,2,3)$ are displayed
on Figure 5. We see that $(m=2\ ,\ n=2,3)$ solutions have largely similar patterns, except for the
additional backbending showing up in the $n=2$ case. The branch structure for $(m=2\ ,\ n=1)$ on the
other hand is drastically different. Indeed it seems quite reminiscent of the corresponding spherically
symmetric $(m=1\ ,\ n=1)$ solutions, pehaps exhibiting a {\it conical fixed point} behaviour too, but the
numerical accuracy is entirely inadequate to decide this, either way.

Our results concerning $m\ge 2$ solutions are here restricted to the concrete construction of $m=2$
solutions. However, our preliminary numerical results indicate that 
most qualitative features remain true for $m=3,~4$. Also, for $m=2$, we
have restricted to $n=1,2,3$ only, which is quite adequate. An interesting remark on $m=2$ solutions
is the existence of the $(m=2\ ,\ n=1)$ confirmed by our results, which we had not found for
the $d=4+0$ dimensional $p=1$ YM model in \cite{Radu:2006gg}, and for which the analytic proof of
existence is also absent. This is easy to understand since our $(m=2\ ,\ n=1)$ solutions 
exist only for that
range of the parameter $\al$ for which all terms in the Lagrangian \re{L} contribute to the action,
and for the limiting range of $\al$ for which only the usual YM term ($p=1$ term) would dominate,
there are no solutions. So there is no special case of our solutions which could 
describe $(m\ge 2\ ,\ n=1)$
bi-azimuthal instantons that are absent in \cite{Radu:2006gg}. 
 
Apart from the above features, we have found a very rich pattern of the zeros of the moduli of
the effective Higgs fields $\chi$ and $\xi$, on the branches parametrised by $\al$. These display
vortex ring structures described in detail in Section 3, similar to the vortex rings discovered
previously~\cite{Kleihaus:2003nj} in the $3+1$ dimensional Yang-Mills--Higgs system.

\bigskip
\noindent
{\bf\large Acknowledgements} \\
This work is carried out in the framework of Enterprise--Ireland Basic Science
Research  Project SC/2003/390 of Enterprise-Ireland. The collaboration with Ya. Shnir
was supported by a Research Enhancement Grant from the Office of
the Dean of Research and Postgraduate Studies of the NUI Maynooth.

\begin{small}

\end{small}
\newpage
\setlength{\unitlength}{1cm}
\begin{picture}(18,8)
\centering
\put(2,0.0){\epsfig{file=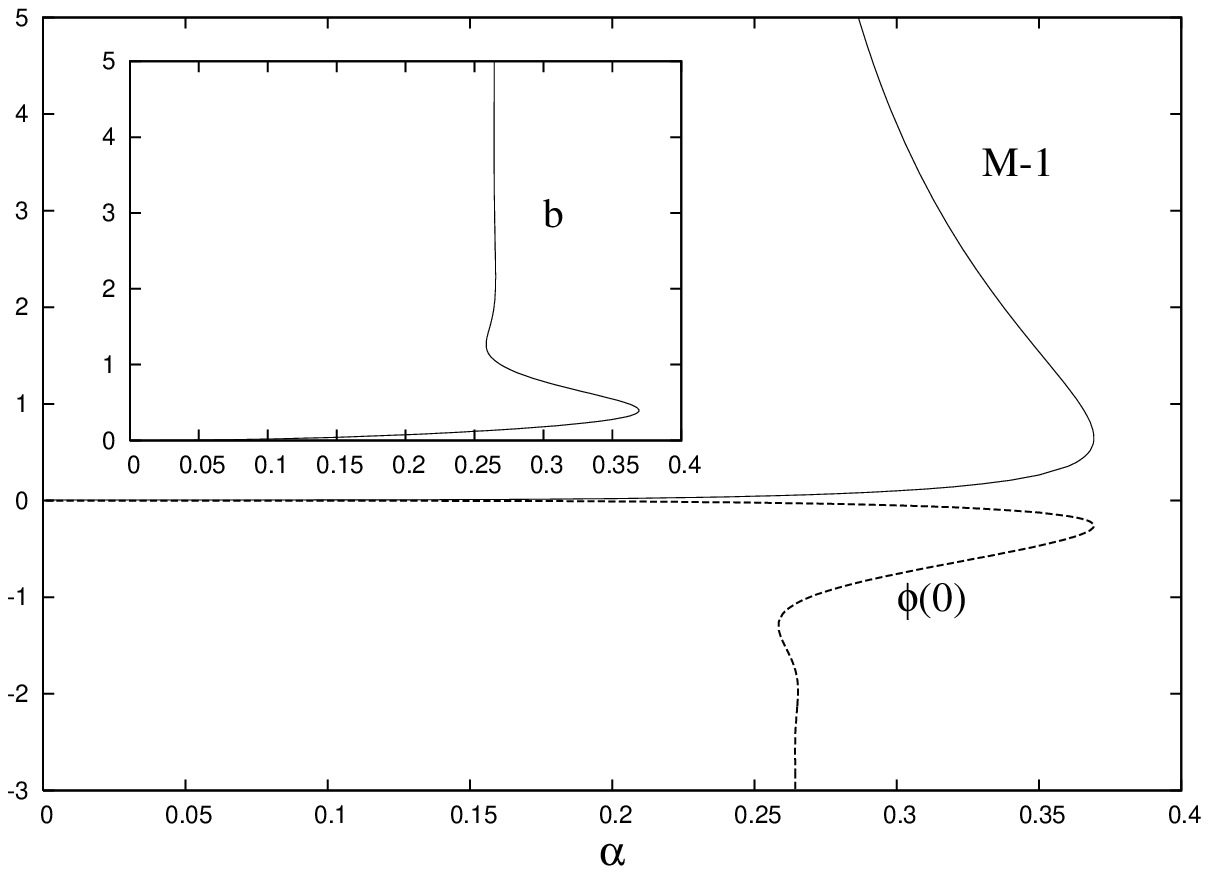,width=13cm}}
\end{picture}
\\
\\
{\small {\bf Figure 1.}
The  mass  $M$, 
 the shooting parameter $b$ and the value of the dilaton at the origin 
$\phi(0)$ are shown as a function of
$\alpha$ for $n=1,~m=1$ spherically symmetric YMd solutions.
(Here and in Figure 2, we use a normalization such that the mass
of $\alpha=0$ self-dual $p=1$ YM solutions is $M=q=n^2$.) }
\\
\\
\\
\\
\setlength{\unitlength}{1cm}
\begin{picture}(19,8)
\centering
\put(2.6,0.0){\epsfig{file=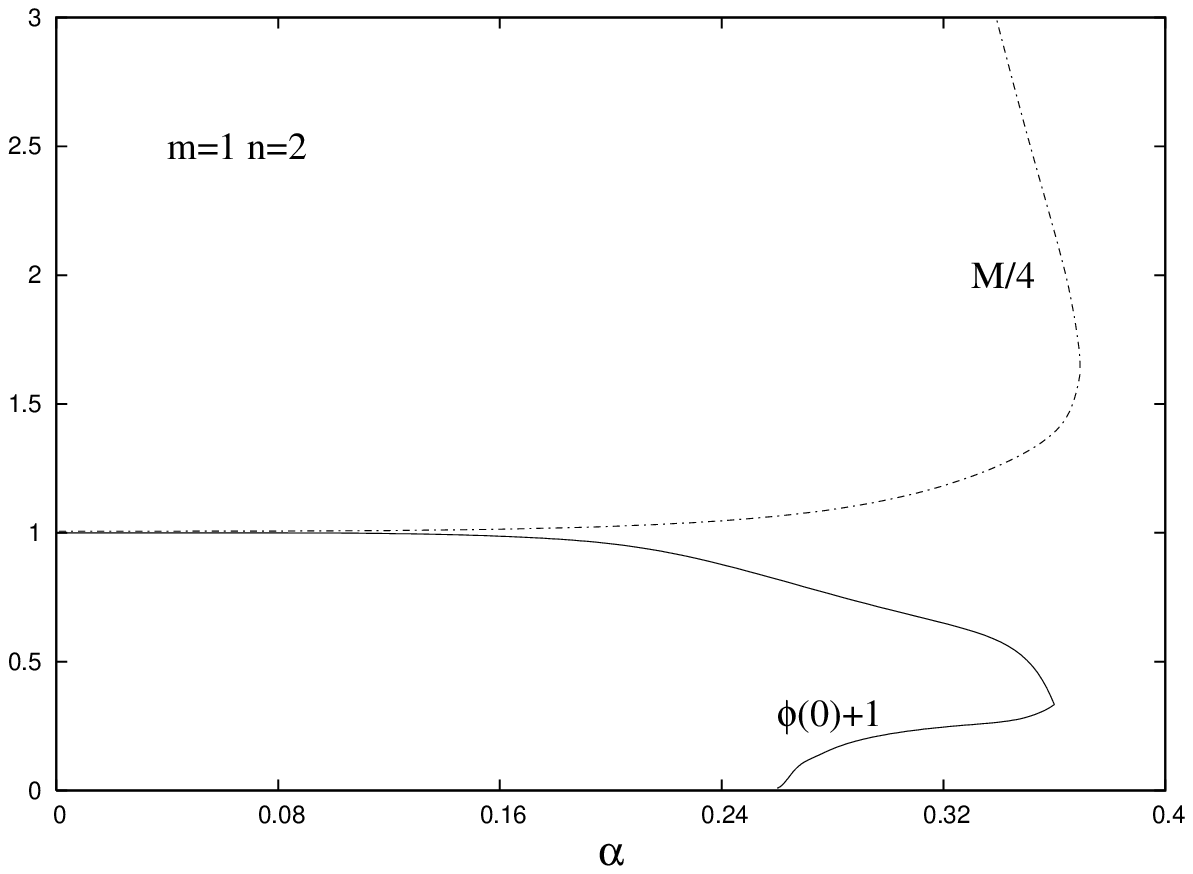,width=13cm}}
\end{picture}
\\
\\
{\small {\bf Figure 2.}
The mass 
 $M$ and the value of the dilaton field at the origin $\phi(0)$
are shown as a function of
$\alpha$ for $n=1,~m=2$ YMd solutions.}
 
\newpage
\setlength{\unitlength}{1cm}
\begin{picture}(18,8)
\centering
\put(2,0.0){\epsfig{file=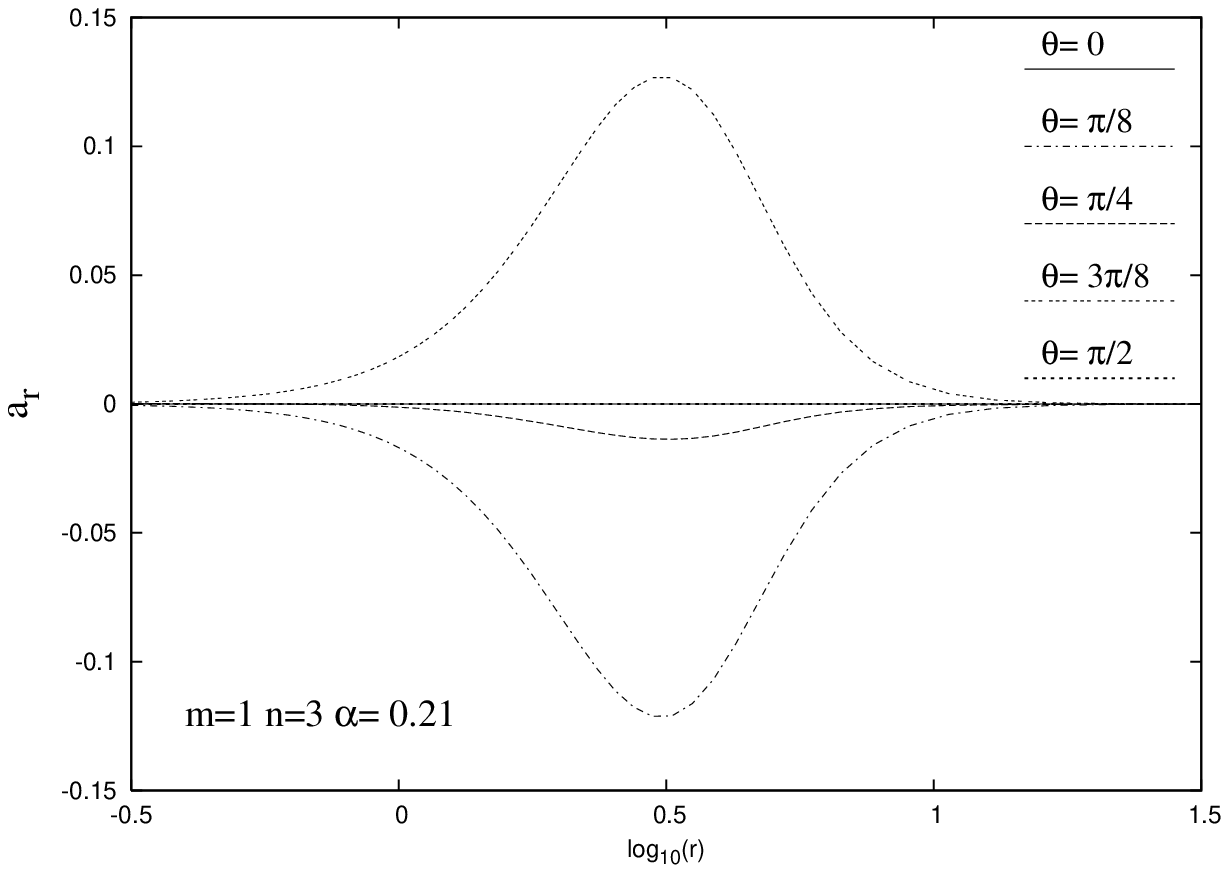,width=13cm}}
\end{picture}
\\
\\
{\small {\bf Figure 3a.}
}
\\
\\
\\
\\
\\
\setlength{\unitlength}{1cm}
\begin{picture}(19,8)
\centering
\put(2.6,0.0){\epsfig{file=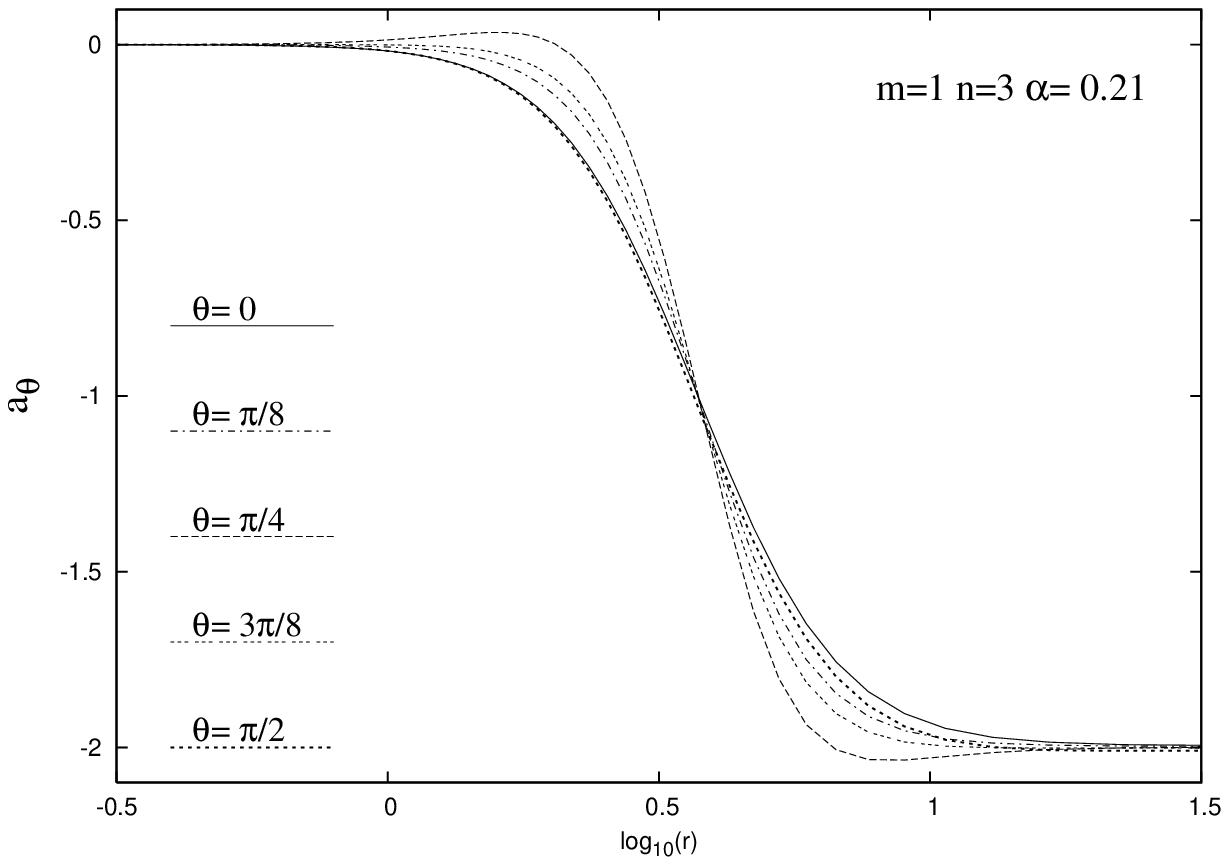,width=13cm}}
\end{picture}
\\
\\
{\small {\bf Figure 3b.} }

\newpage
\setlength{\unitlength}{1cm}
\begin{picture}(18,8)
\centering
\put(2,0.0){\epsfig{file=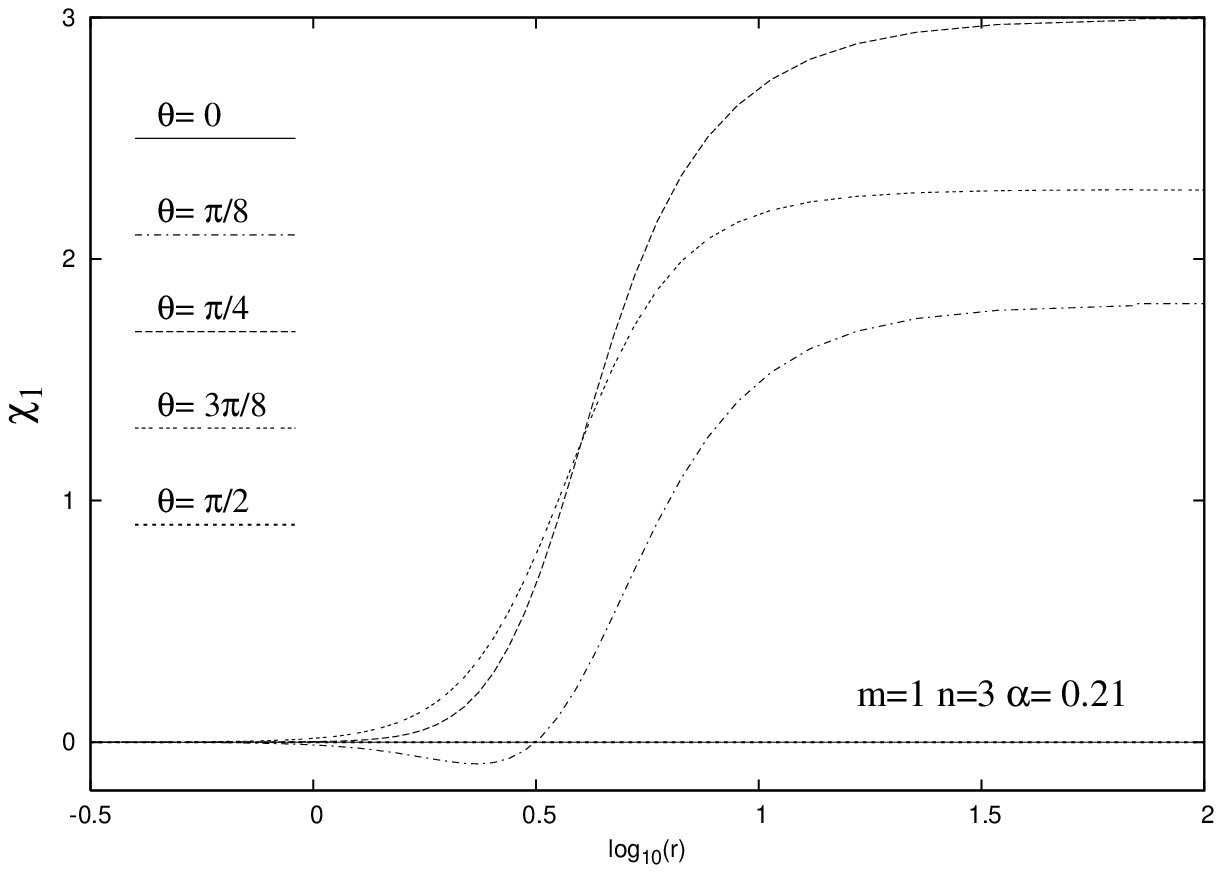,width=13cm}}
\end{picture}
\\
\\
{\small {\bf Figure 3c.} }
\\
\\
\\
\\
\\
\setlength{\unitlength}{1cm}
\begin{picture}(19,8)
\centering
\put(2.6,0.0){\epsfig{file=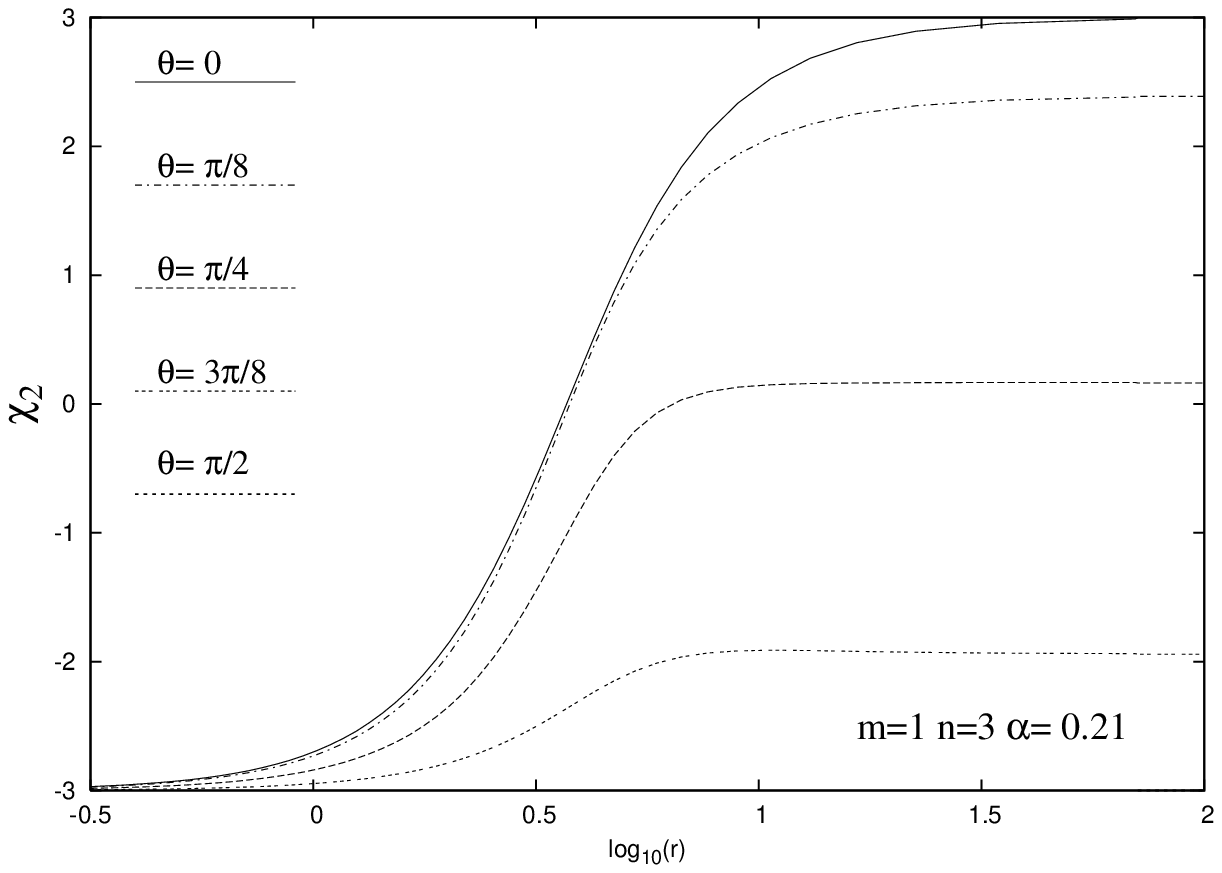,width=13cm}}
\end{picture}
\\
\\
{\small {\bf Figure 3d.} }

\newpage
\setlength{\unitlength}{1cm}
\begin{picture}(18,8)
\centering
\put(2,0.0){\epsfig{file=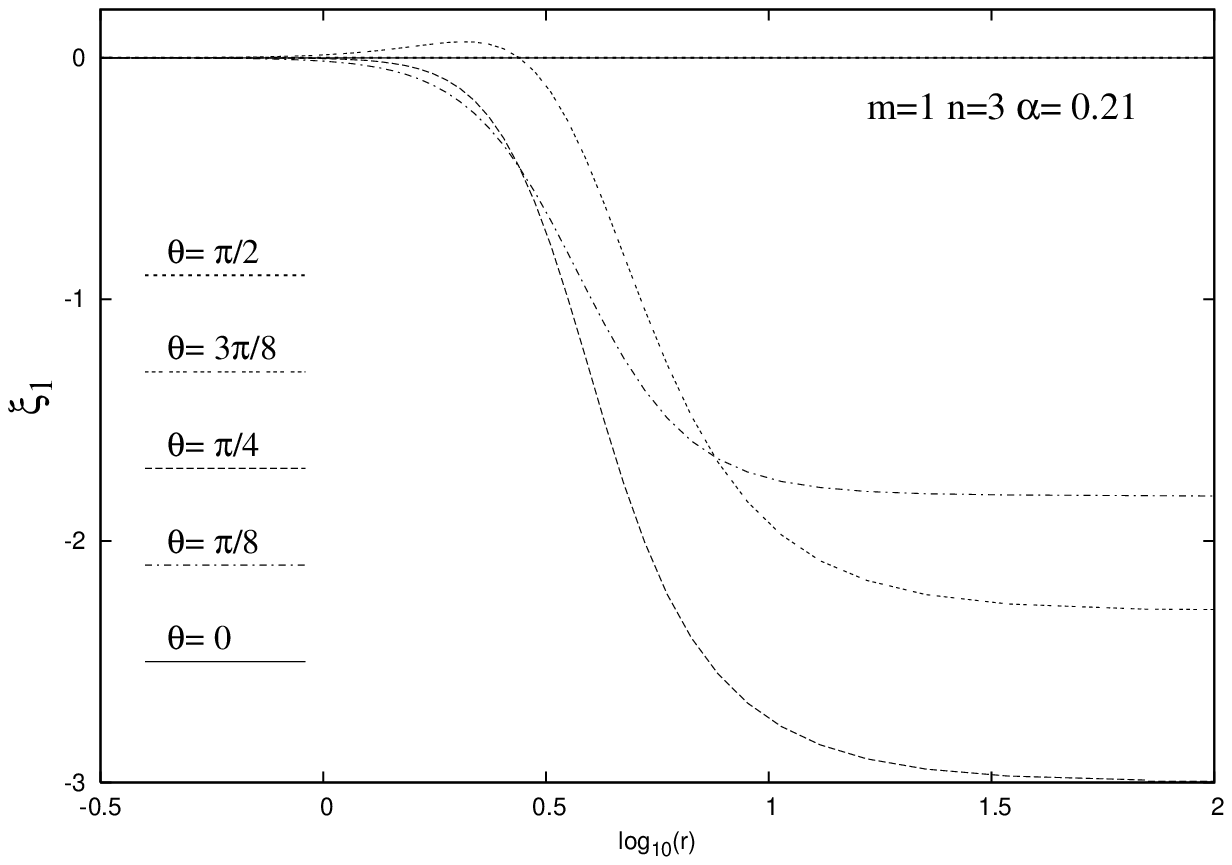,width=13cm}}
\end{picture}
\\
\\
{\small {\bf Figure 3e.} }
\\
\\
\\
\\
\\
\setlength{\unitlength}{1cm}
\begin{picture}(19,8)
\centering
\put(2.6,0.0){\epsfig{file=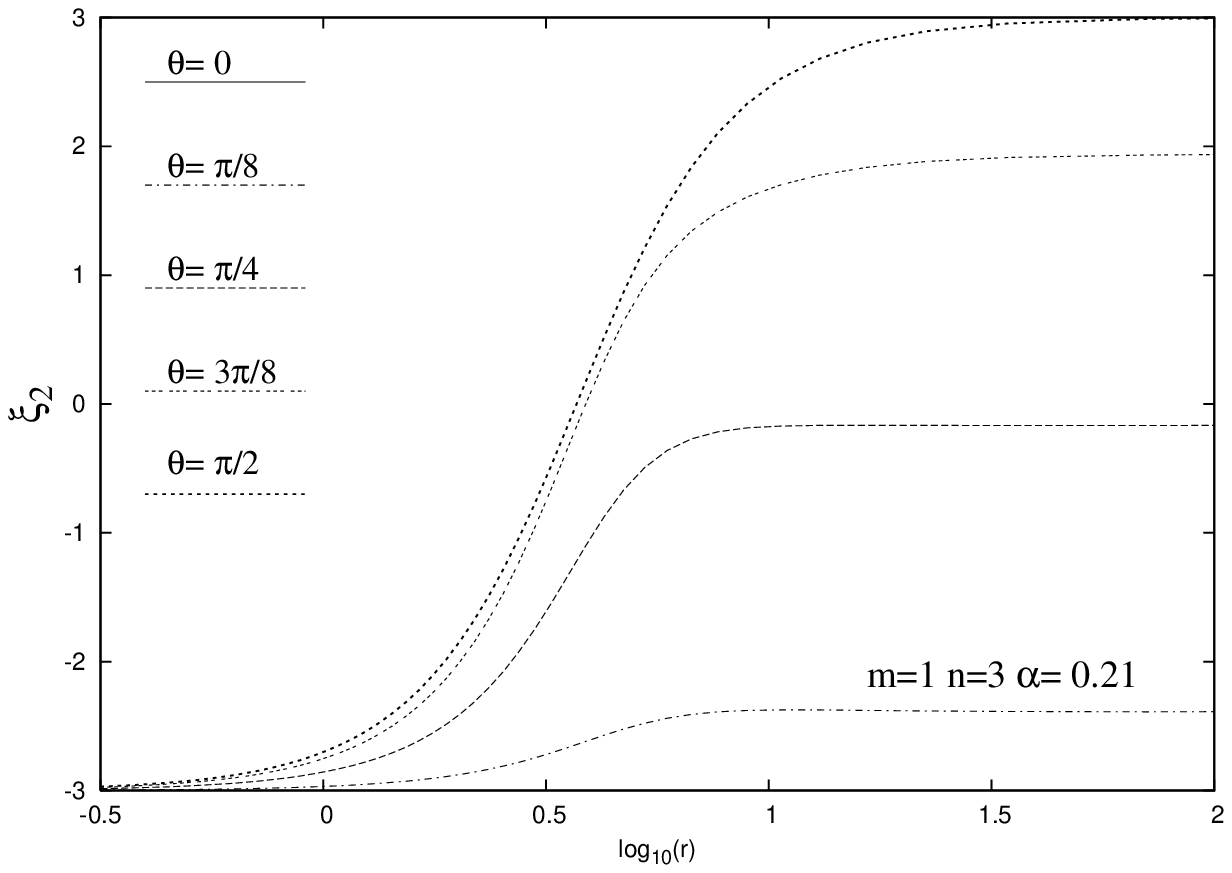,width=13cm}}
\end{picture}
\\
\\
{\small {\bf Figure 3f.} }

 \newpage
\setlength{\unitlength}{1cm}
\begin{picture}(19,8)
\centering
\put(2.,0.0){\epsfig{file=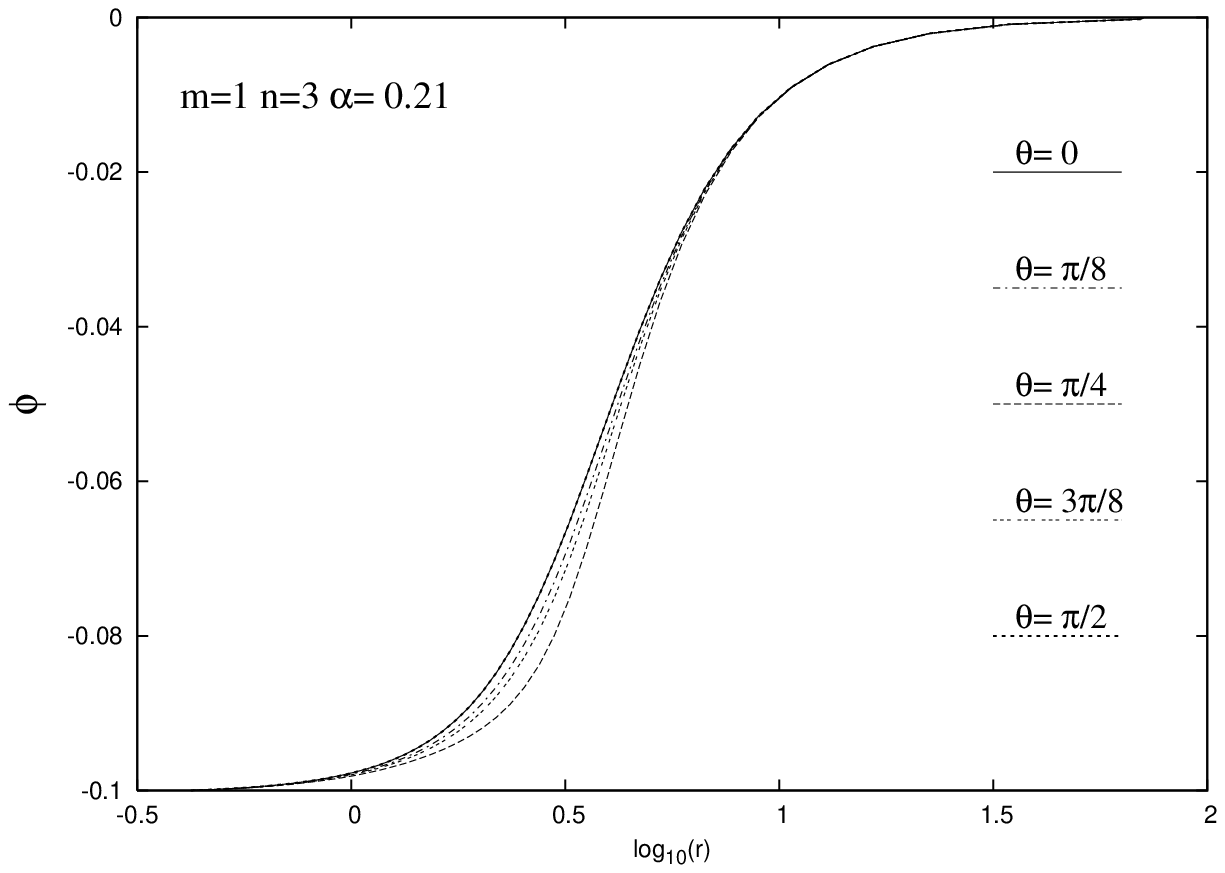,width=13cm}}
\end{picture}
\\
\\
{\small {\bf Figure 3g.}}
\\
\\
\\
\\
\\
\setlength{\unitlength}{1cm}
\begin{picture}(19,8)
\centering
\put(2.6,0.0){\epsfig{file=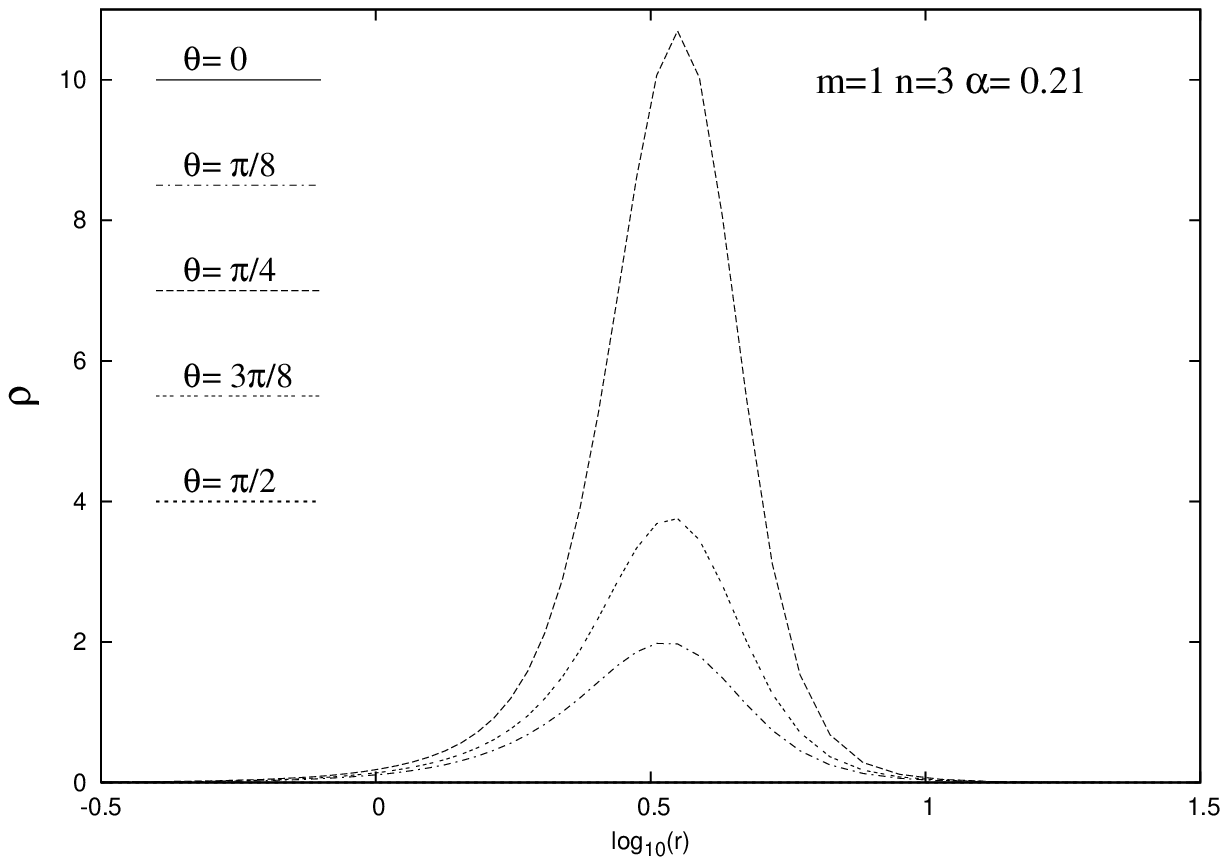,width=13cm }}
\end{picture}
\\
\\
{\small {\bf Figure 3h.}
The YM gauge functions, the dilaton field and the 
topological charge density $\rho$ 
are shown as a function of the radial coordinate
$r$  for a typical  $m=1,~n=3$
YMd solutions with $\alpha= 0.21$. }

 \newpage
\setlength{\unitlength}{1cm}
\begin{picture}(1,1)
\centering
\put(2.,3.2){\epsfig{file=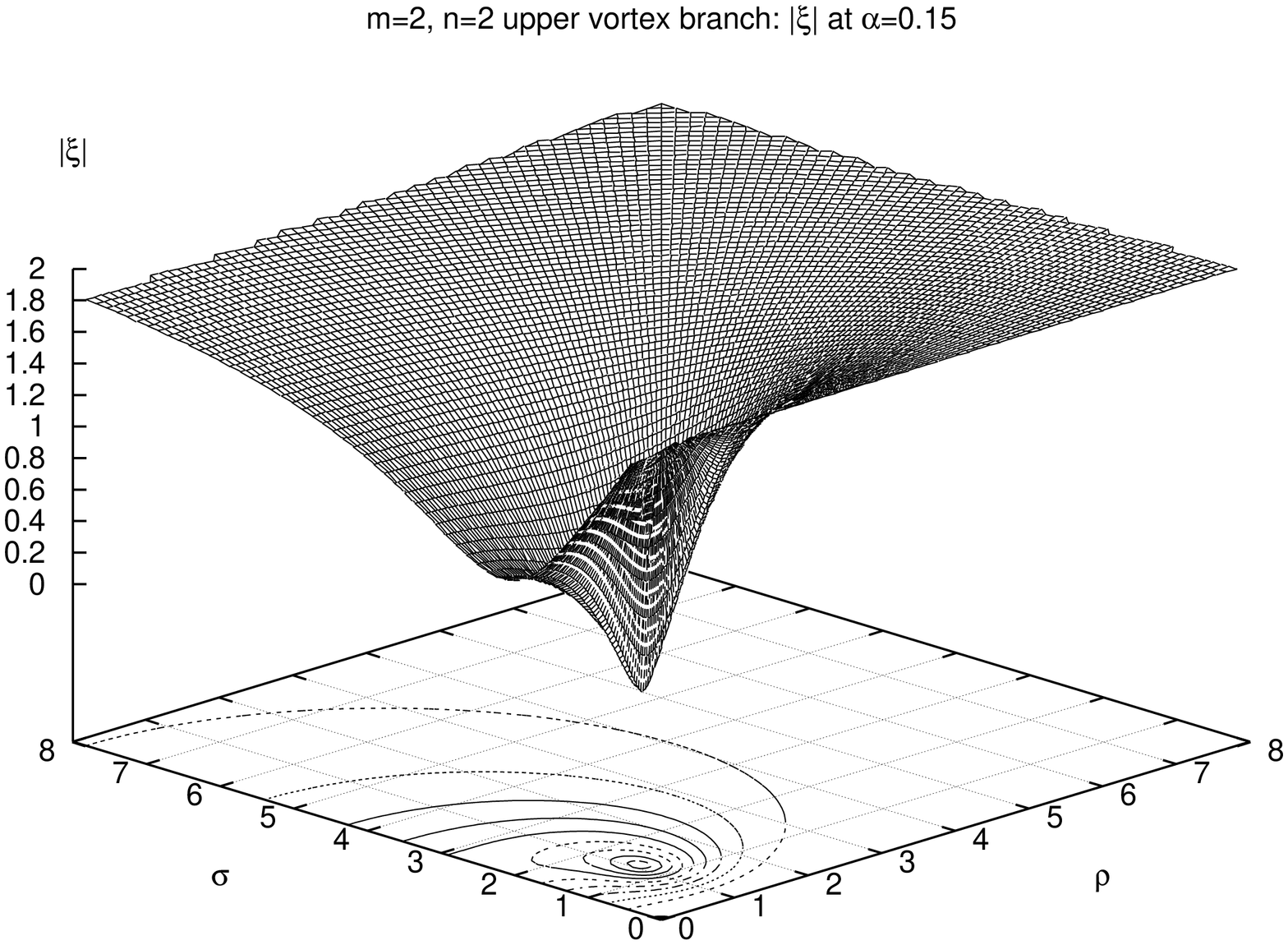,width=11cm, angle =-90}}
\end{picture}
\vspace{8.cm}
\\
{\small {\bf Figure 4a.}}
\\
\\
\\
\setlength{\unitlength}{1cm}
\begin{picture}(19,8)
\centering
\put(2.6,9.0){\epsfig{file=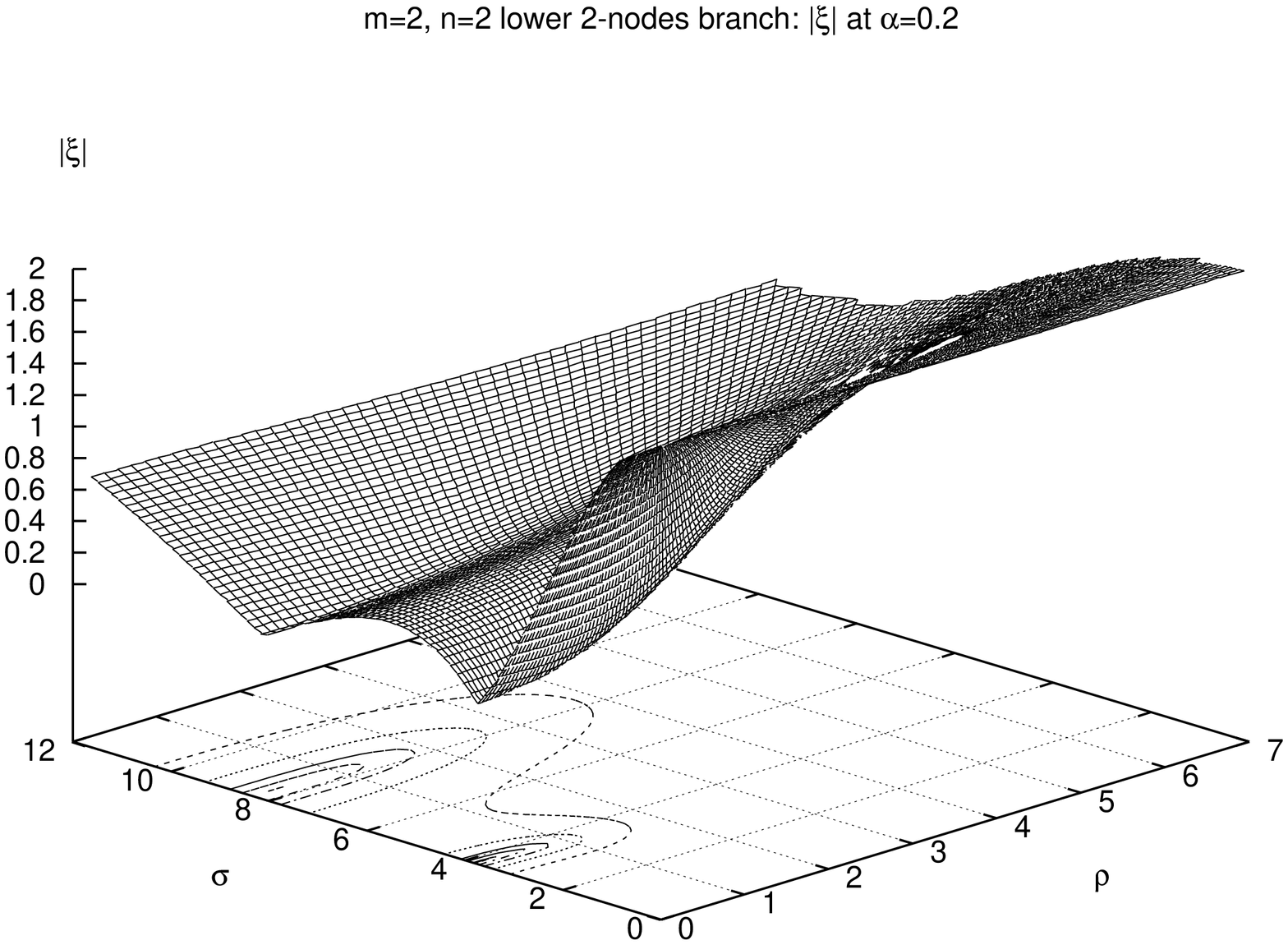,width=11cm , angle =-90}}
\end{picture}
\\
\\
\\
{\small {\bf Figure 4b.} }
\newpage
\setlength{\unitlength}{1cm}
\begin{picture}(1,1)
\centering
\put(2.,3.2){\epsfig{file=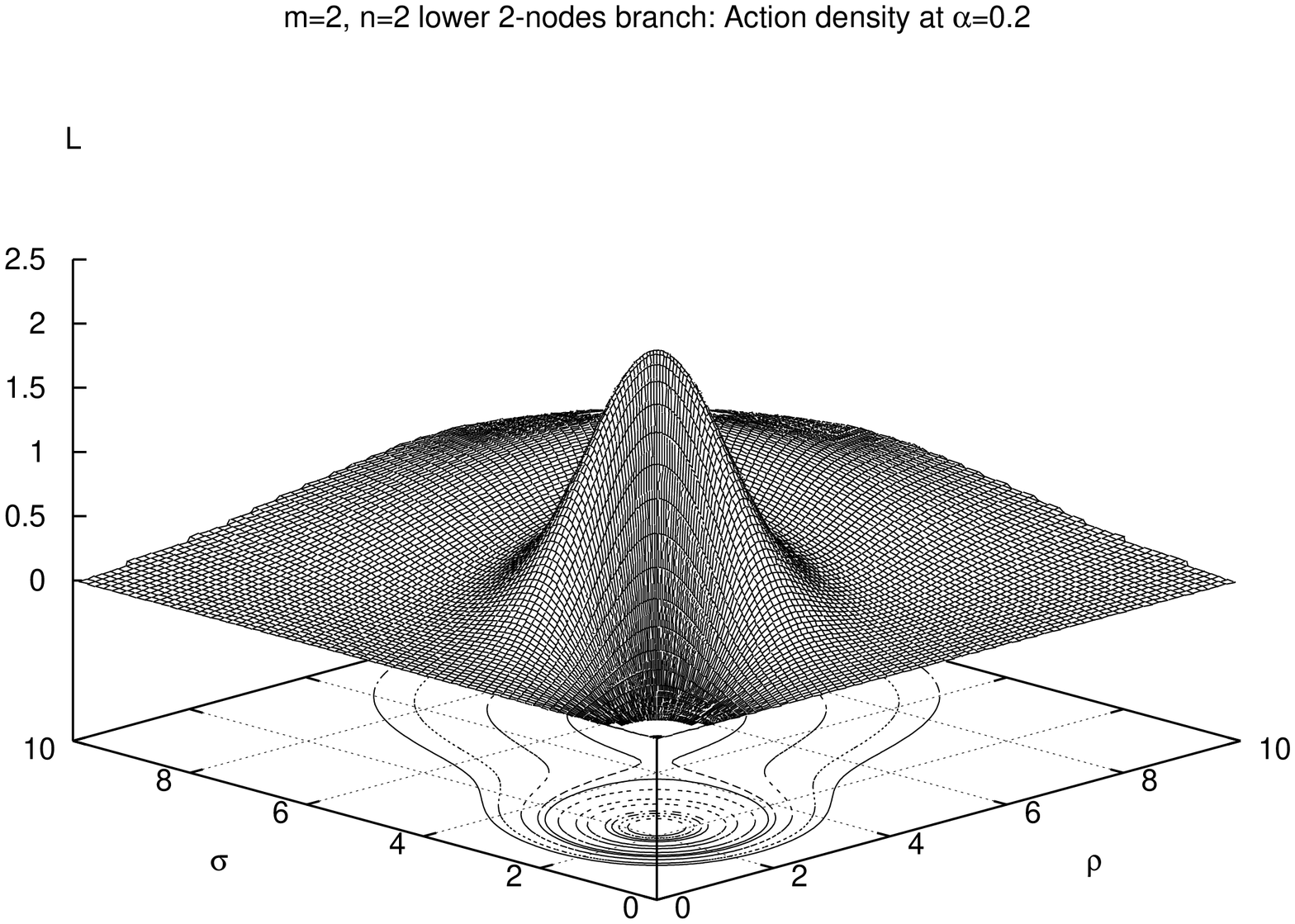,width=11cm, angle =-90}}
\end{picture}
\vspace{8.cm}
\\
{\small {\bf Figure 4c.} 
\\
\\
\setlength{\unitlength}{1cm}
\begin{picture}(19,8)
\centering
\put(2.6,9.5){\epsfig{file=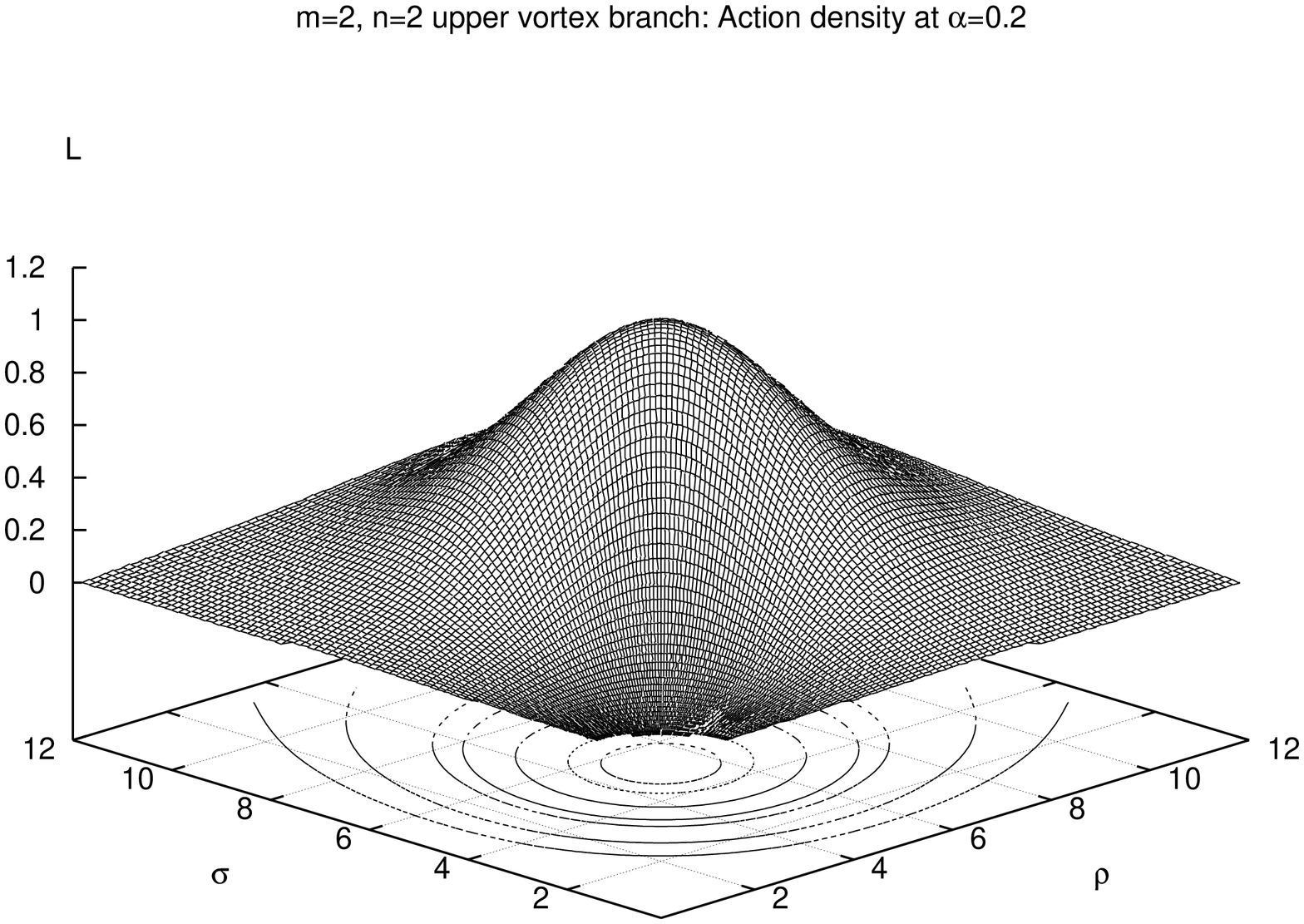,width=11cm , angle =-90}}
\end{picture}
\\
\\
{\small {\bf Figure 4d.} }
 The modulus of the effective Higgs field $\xi$ is shown for the 
uper branch $m=2, n=2$ solutions at $\alpha = 0.15$ (vortex, Figure 4a) and $\alpha = 0.20$ 
lower energy branch solution (double node, Figure 4b) as functions of the coordinates $\rho$ 
and $\sigma$. The action density distributions of these $m=2, n=2$ solutions at 
$\alpha = 0.20$ are also shown on the lower branch (Figure 4c) 
and on the upper branch (Figure 4d),  respectively. 
}
\newpage
\setlength{\unitlength}{1cm}
\begin{picture}(18,8)
\centering
\put(2,0.0){\epsfig{file=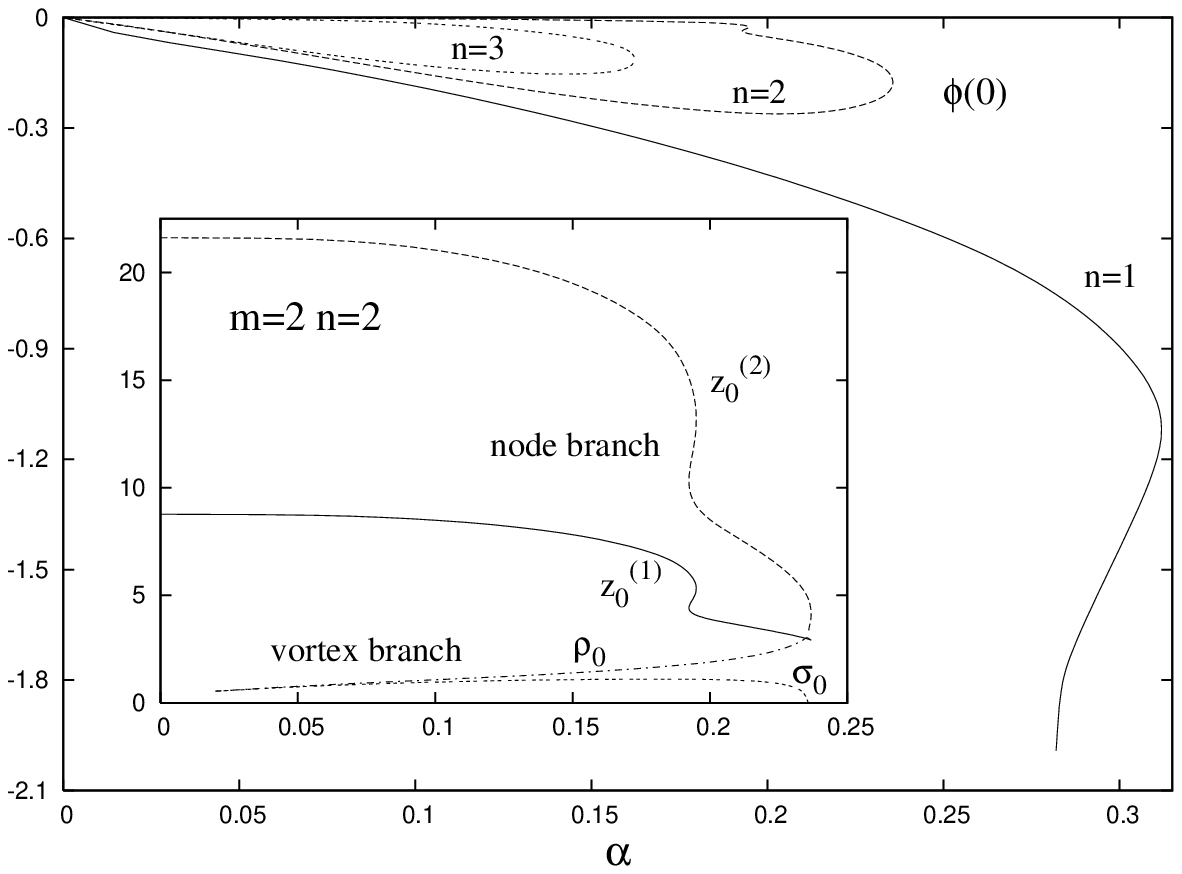,width=13cm}}
\end{picture}
\\
{\small {\bf Figure 5a.}
}
\\
\\
\\
\\
\\
\setlength{\unitlength}{1cm}
\begin{picture}(19,8)
\centering
\put(2.1,0.0){\epsfig{file=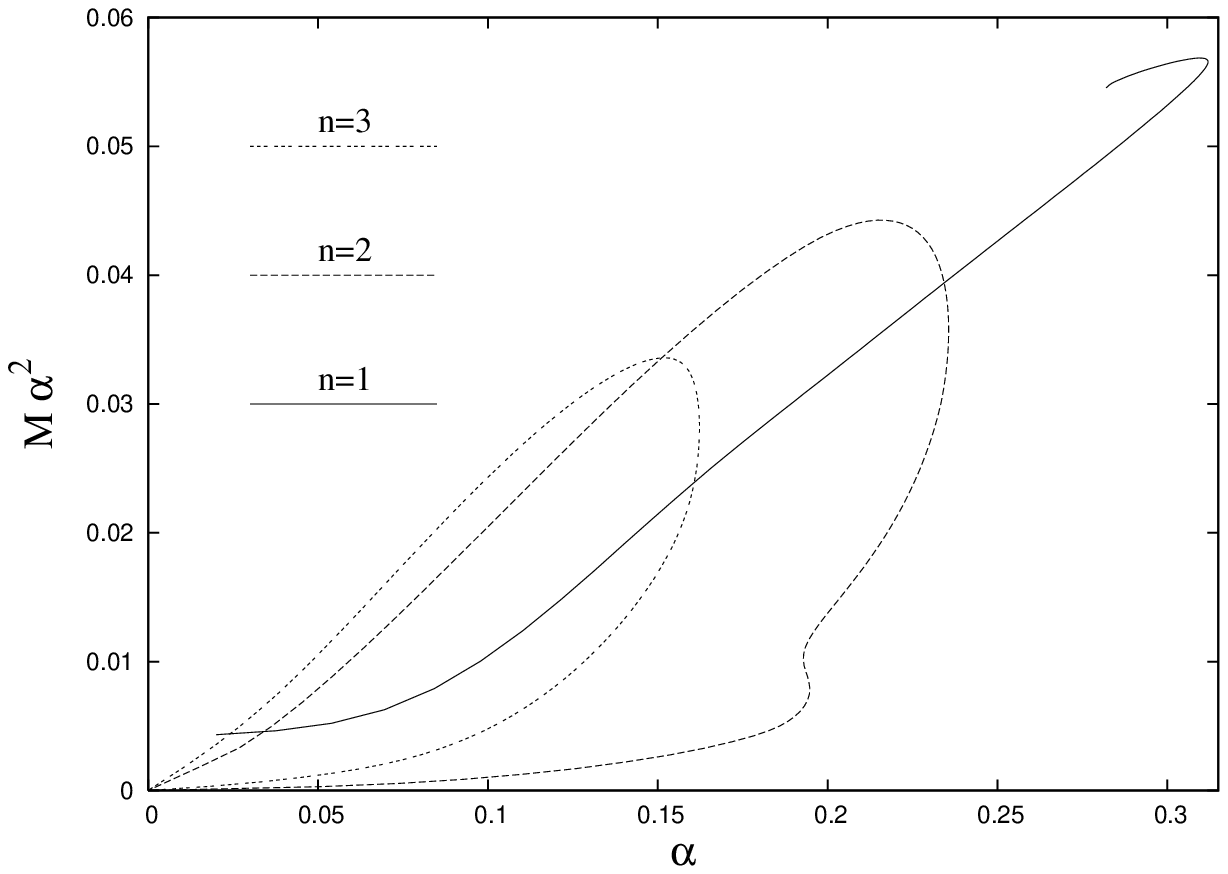,width=13.5cm}}
\end{picture}
\\
\\
\\
{\small {\bf Figure 5b.}
The values of the dilaton function $\phi$ at the origin (Figure 5a) and the unrescaled mass
$M\alpha^2$  (Figure 5b) of the configuration with $m=2,~n=1,2,3$ are shown as functions of the 
effective coupling constant $\alpha$.
The location of the vortices as given by $(\rho_0,\sigma_0)$   
and the position of nodes $(z_0^{(1)},~z_0^{(2)})$ 
is also presented in Figure 5a for $m=2,~n=2$ solutions.
}

\newpage
\setlength{\unitlength}{1cm}
\begin{picture}(18,8)
\centering
\put(2,0.0){\epsfig{file=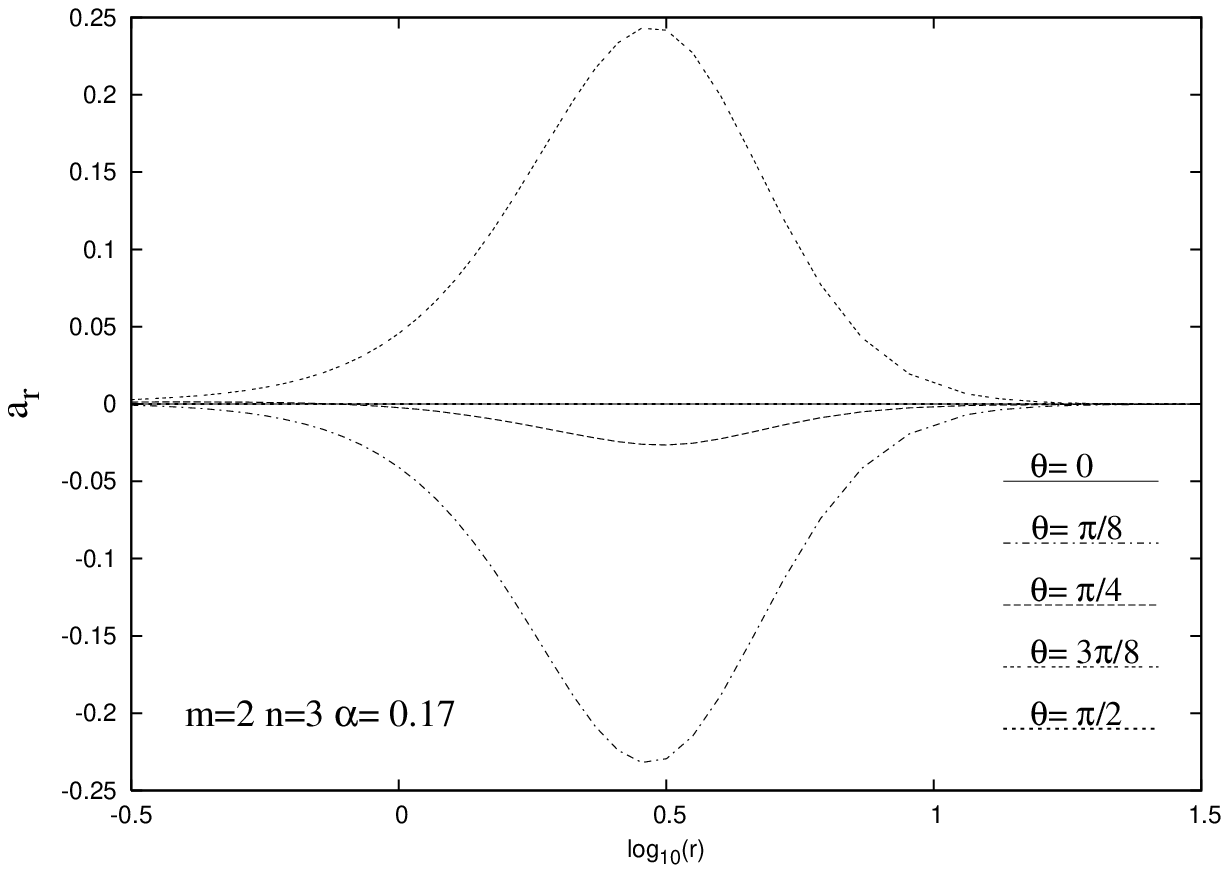,width=13cm}}
\end{picture}
\\
\\
{\small {\bf Figure 6a.}
}
\\
\\
\\
\\
\\
\setlength{\unitlength}{1cm}
\begin{picture}(19,8)
\centering
\put(2.6,0.0){\epsfig{file=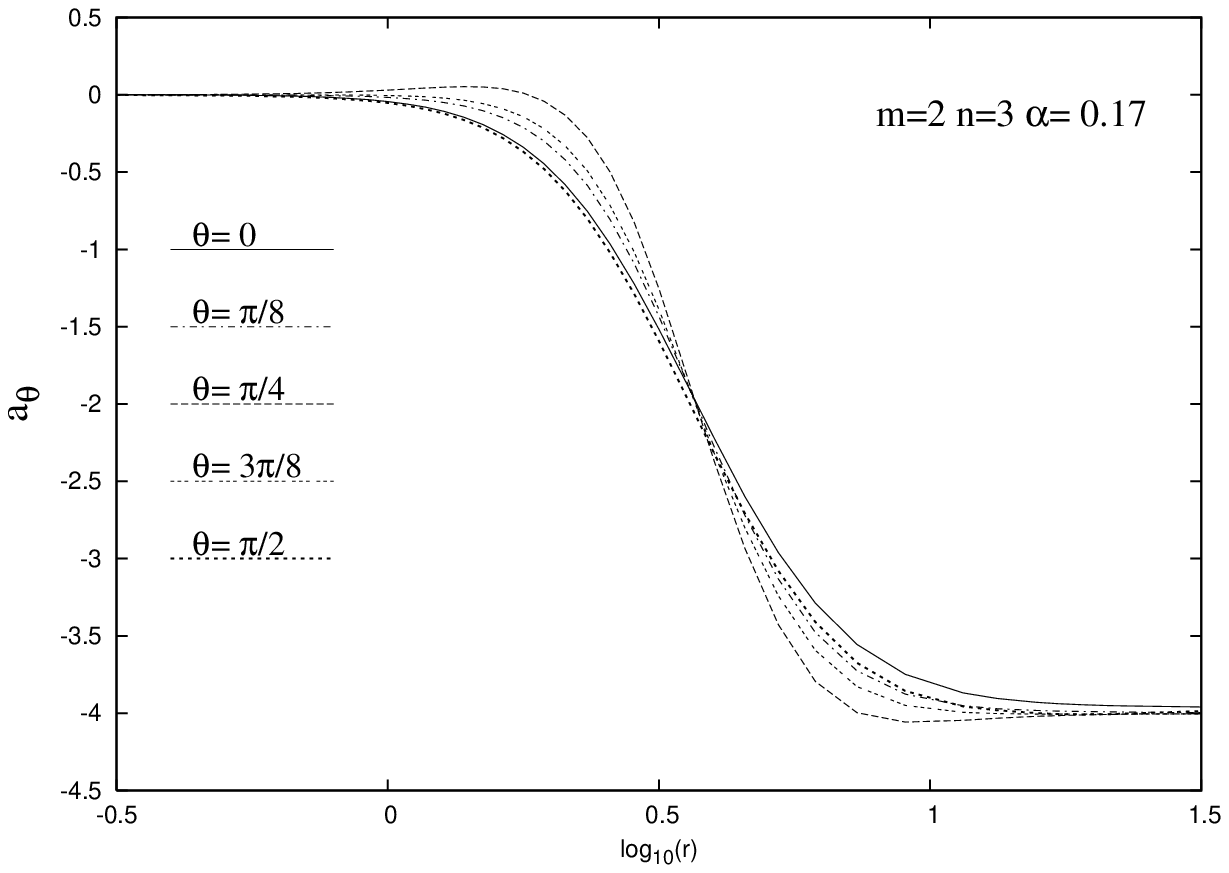,width=13cm}}
\end{picture}
\\
\\
{\small {\bf Figure 6b.} }
\newpage
\setlength{\unitlength}{1cm}
\begin{picture}(18,8)
\centering
\put(2,0.0){\epsfig{file=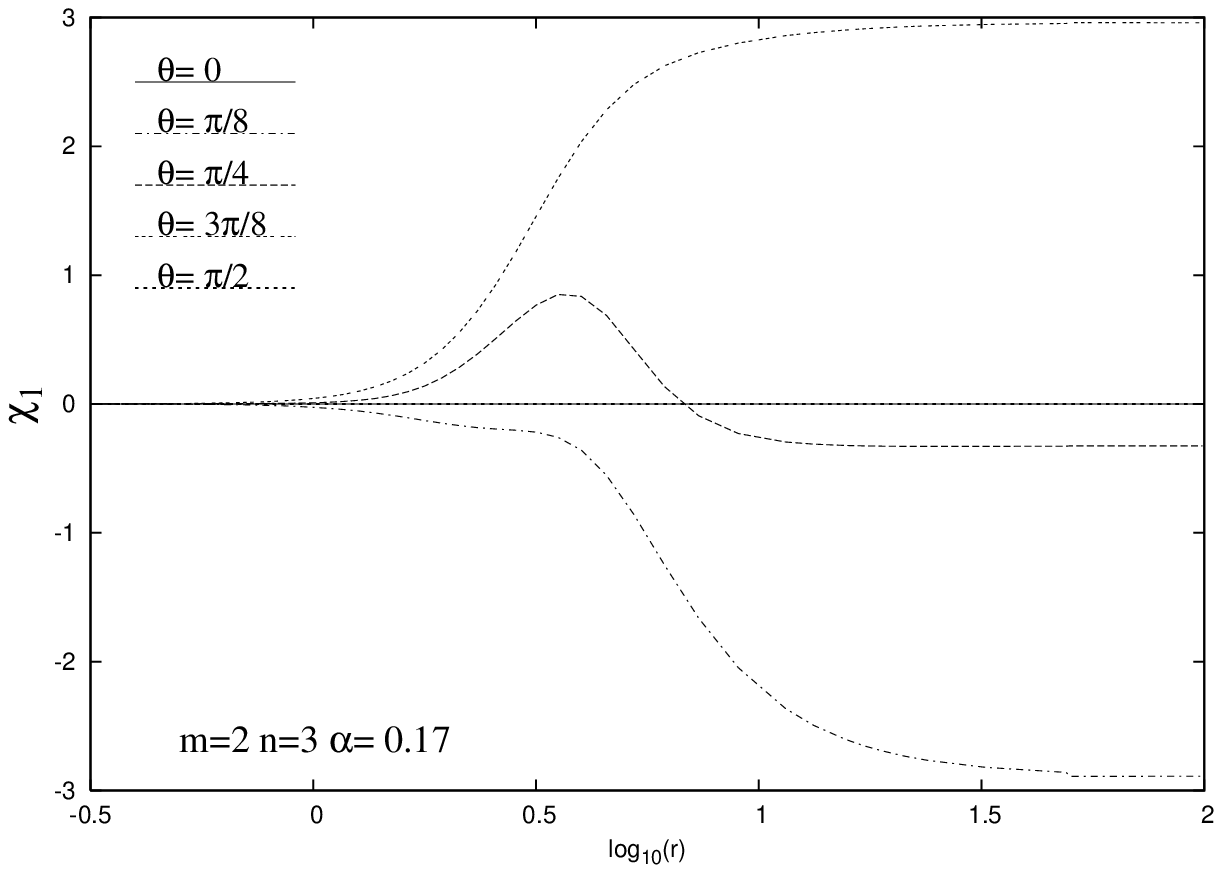,width=13cm}}
\end{picture}
\\
\\
{\small {\bf Figure 6c.}
}
\\
\\
\\
\\
\\
\setlength{\unitlength}{1cm}
\begin{picture}(19,8)
\centering
\put(2.6,0.0){\epsfig{file=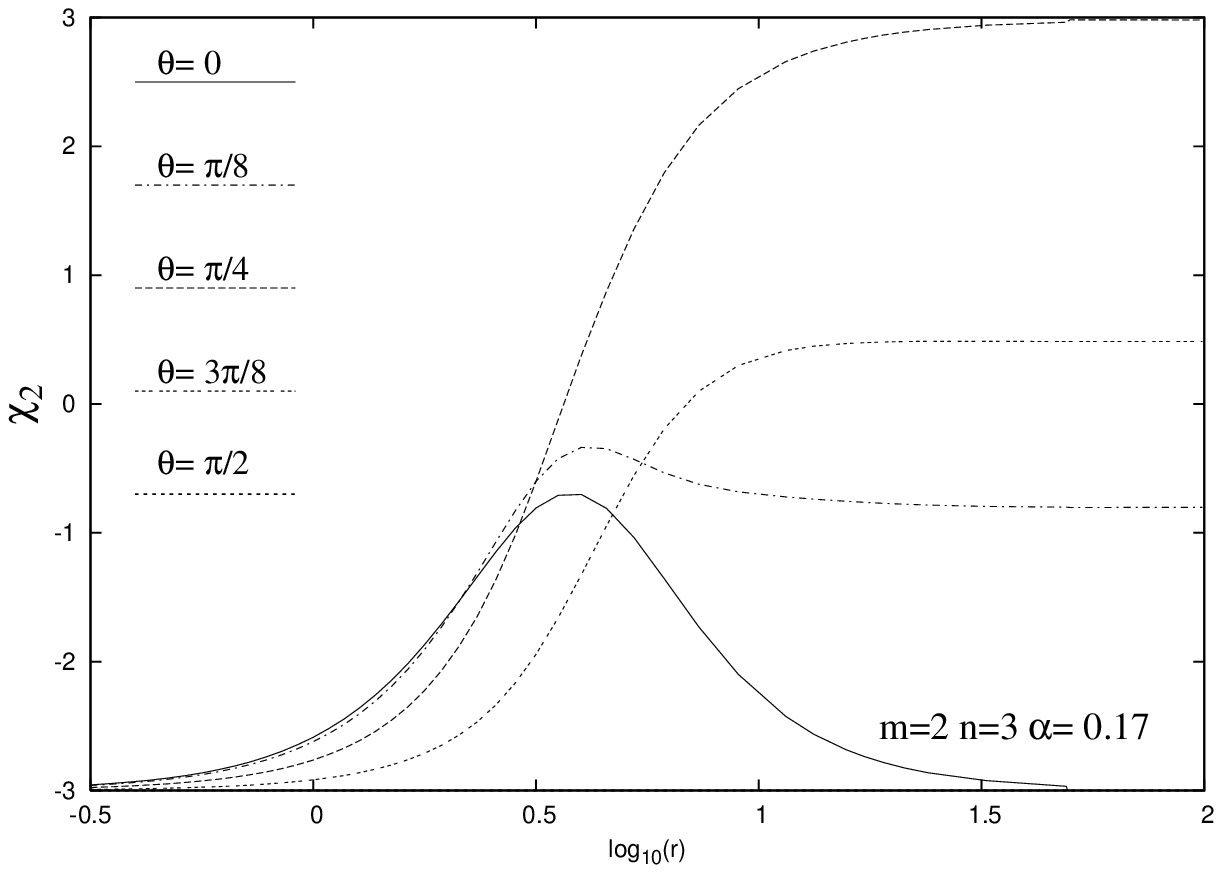,width=13cm}}
\end{picture}
\\
\\
{\small {\bf Figure 6d.} }
\newpage
\setlength{\unitlength}{1cm}
\begin{picture}(18,8)
\centering
\put(2,0.0){\epsfig{file=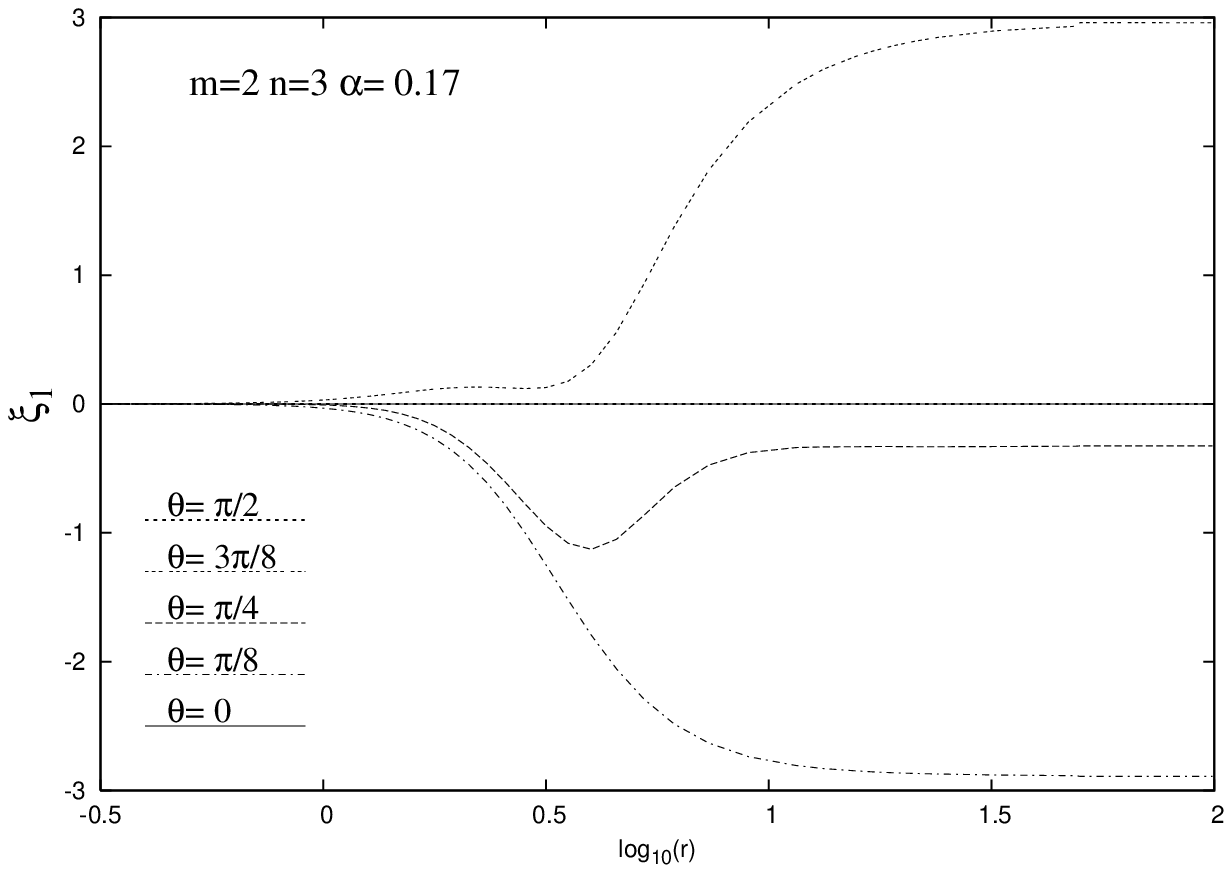,width=13cm}}
\end{picture}
\\
\\
{\small {\bf Figure 6e.}
}
\\
\\
\\
\\
\\
\setlength{\unitlength}{1cm}
\begin{picture}(19,8)
\centering
\put(2.6,0.0){\epsfig{file=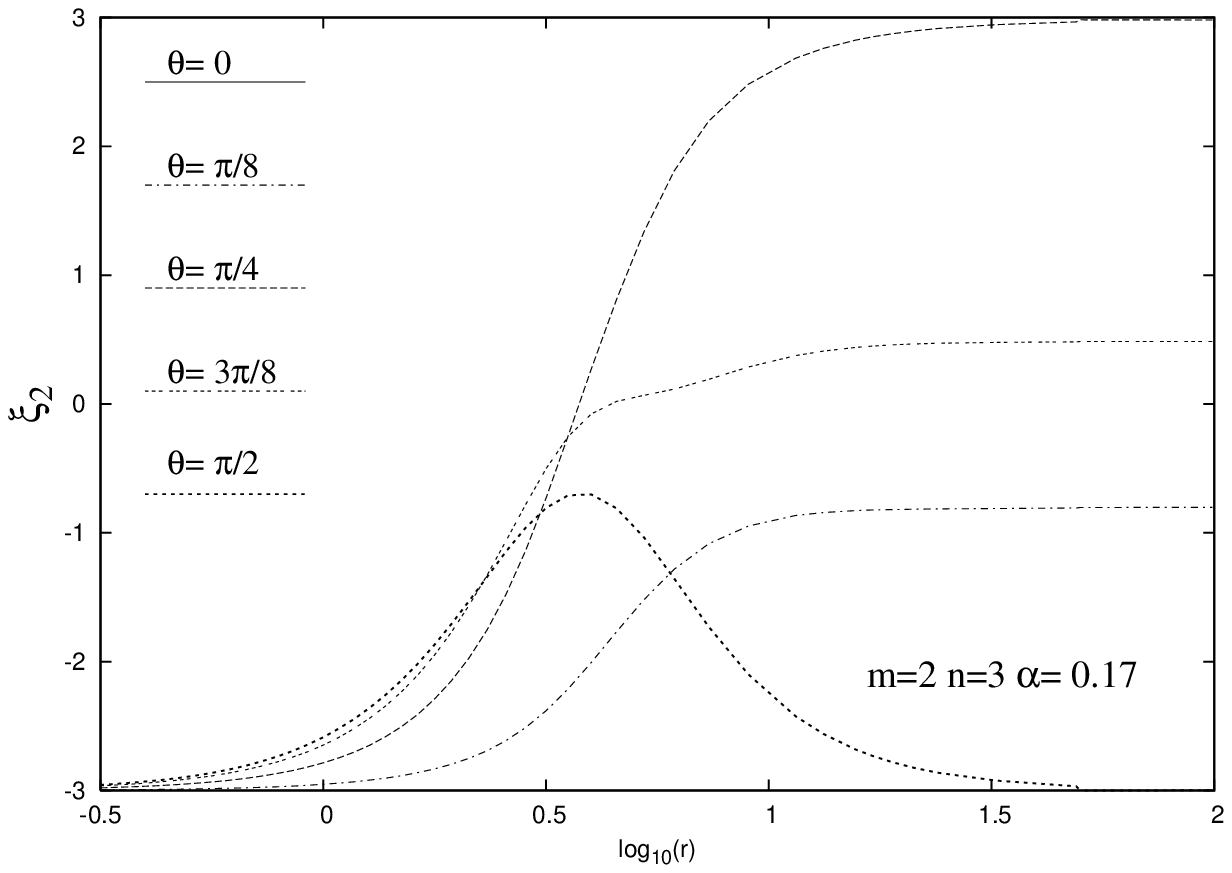,width=13cm}}
\end{picture}
\\
\\
{\small {\bf Figure 6f.} }
\newpage
\setlength{\unitlength}{1cm}
\begin{picture}(18,8)
\centering
\put(2,0.0){\epsfig{file=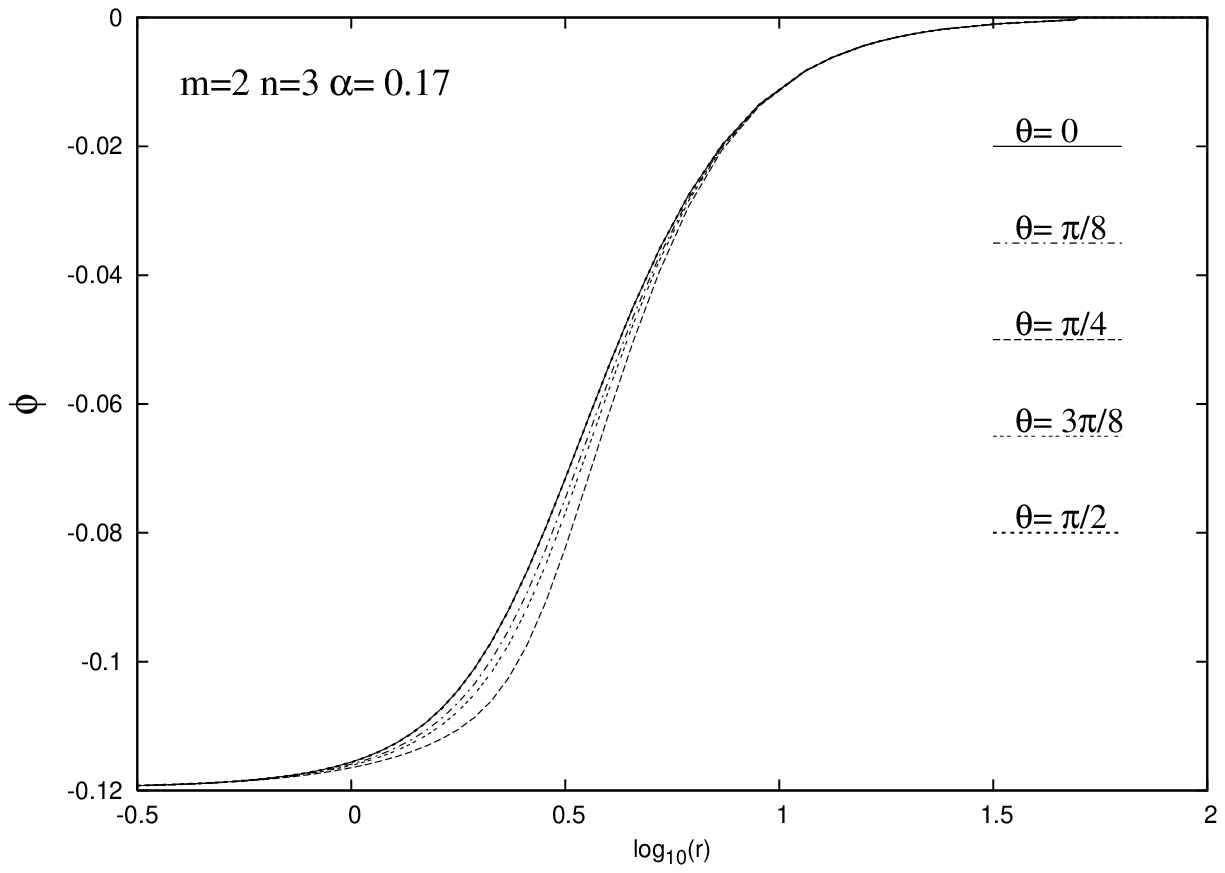,width=13cm}}
\end{picture}
\\
\\
{\small {\bf Figure 6g.} 
The YM gauge functions and the dilaton field  
are shown as a function of the radial coordinate
$r$  for a typical  $m=2,~n=3$
YMd solutions with $\alpha= 0.17$. 
}

\end{document}